\documentclass{SciPost}

\hypersetup{
    colorlinks,
    linkcolor={red!50!black},
    citecolor={blue!50!black},
    urlcolor={blue!80!black}
}

\usepackage{nicefrac}
\usepackage{amsmath}
\usepackage{physics}
\usepackage{float}
\usepackage{subcaption}
\usepackage[bitstream-charter]{mathdesign}
\usepackage{csquotes}
\usepackage[normalem]{ulem}

\DeclareSymbolFont{usualmathcal}{OMS}{cmsy}{m}{n}
\DeclareSymbolFontAlphabet{\mathcal}{usualmathcal}

\fancypagestyle{SPstyle}{
\fancyhf{}
\lhead{\colorbox{scipostblue}{\bf \color{white} ~SciPost Physics Core }}
\rhead{{\bf \color{scipostdeepblue} ~Submission }}

\fancyfoot[C]{\textbf{\thepage}}
}

\newcommand{\realpart}{\mathrm{Re}}     
\newcommand{\Qu}{\mathrm{Q}}            

\newcommand{\mb}{\mathbf}

\def\op#1{\mb{#1}}

\newcommand{\oW}{\op{W}}

\newcommand{\oU}{\op{U}}

\newcommand{\oS}{\op{S}}

\newcommand{\Sa}{\oS^{\alpha}}
\newcommand{\Sb}{\oS^{\beta}}

\newcommand{\oV}{\op{V}}
\newcommand{\oVt}{\tilde{\op{V}}}

\def\exp#1{\mathrm{e}^{#1}}

\newcommand{\JQ}{J_{\Qu}}
\newcommand{\JL}{J_{\mathrm{L}}}

\newlength{\textarrow}
\def\trace#1{\mathrm{Tr}\left\{#1\right\}}

\usepackage{tikz} 

\usepackage{tikz} 

\begin{document}

\pagestyle{SPstyle}

\begin{center}{\Large \textbf{\color{scipostdeepblue}{
Dynamical mean-field theory for dense spin systems at finite temperature\\
}}}\end{center}

\begin{center}\textbf{
Przemysław Bieniek\textsuperscript{1$\star$},
Timo Gräßer\textsuperscript{2} and
Götz S. Uhrig\textsuperscript{1$\dagger$}
}\end{center}

\begin{center}
{\bf 1} Condensed Matter Theory, TU Dortmund University, \\Otto-Hahn-Str.\ 4, 44227 Dortmund, Germany
\\
{\bf 2} Institute of Molecular Physical Science, ETH Zürich,\\ Vladimir-Prelog-Weg 1-5/10, 8093 Zürich, Switzerland
\\[\baselineskip]
$\star$ \href{mailto:przemyslaw.bieniek@tu-dortmund.de}{\small przemyslaw.bieniek@tu-dortmund.de}\,,\quad
$\dagger$ \href{mailto:goetz.uhrig@tu-dortmund.de}{\small goetz.uhrig@tu-dortmund.de}
\end{center}

\section*{\color{scipostdeepblue}{Abstract}}
\textbf{\boldmath{%
In recent years, a method for computing spin dynamics at infinite temperature (spinDMFT) was developed. It utilizes the ideas of dynamical mean-field theory for fermions: single-site approximation and a self-consistency condition to approximate time-dependent spin correlations. In this work, we develop a crucial extension of the method to systems at finite temperature, able to compute imaginary-time correlations and thermodynamical quantities. We benchmark the method by comparison to results in finite-size systems, obtaining very good agreement with correlations in a random-coupling system, good agreement for a ferromagnetic system and large discrepancies in the case of an antiferromagnet. We note the appearance of ferromagnetic order in the method. We discuss possible extensions and potential applications of the approach.
}}

\vspace{\baselineskip}

\noindent\textcolor{white!90!black}{%
\fbox{\parbox{0.975\linewidth}{%
\textcolor{white!40!black}{\begin{tabular}{lr}%
  \begin{minipage}{0.6\textwidth}%
    {\small Copyright attribution to authors. \newline
    This work is a submission to SciPost Physics Core. \newline
    License information to appear upon publication. \newline
    Publication information to appear upon publication.}
  \end{minipage} & \begin{minipage}{0.4\textwidth}
    {\small Received Date \newline Accepted Date \newline Published Date}%
  \end{minipage}
\end{tabular}}
}}
}


\vspace{10pt}
\noindent\rule{\textwidth}{1pt}
\tableofcontents
\noindent\rule{\textwidth}{1pt}
\vspace{10pt}


\section{Introduction}
\label{sec:intro}
The study of spin systems is of great interest for both practical applications and theoretical investigation. Materials with relevant spin degrees of freedom are promising candidates for quantum computing \cite{Loss1998,Trauzettel2007}, new methods of memory storage and new electronic devices \cite{Jungwirth2016,Jungfleisch2018,Baltz2018}. Nuclear magnetic resonance (NMR) is another important application, using nuclear spins to reveal the structure of matter \cite{Haeberlen1976,Slichter1996,Levitt2005}. Recent theoretical developments revealed new exciting fields in the research of spin ensembles, in particular regarding new kinds of order -- spin-split antiferromagnetism in systems of reduced symmetry (altermagnets) \cite{Smejkal2022_1, Smejkal2022_2} and topological order \cite{Castelnovo2007, Balents2010}.

The description of dynamics resulting from spin models remains a formidable problem. A wide variety of techniques applicable to spin systems has been developed. The primary challenge is the exponential scaling of the size of the Hilbert space with the number of spins in the system. For this reason, calculations by exact diagonalization (ED) of the Hamiltonian are possible only for small system sizes. Larger systems can be considered when focusing on the low-energy states with the Lanczos algorithm. The Chebyshev expansion technique (CET) can improve upon the efficiency of ED at high temperatures \cite{Tal-Ezer1984, Kosloff1994, Weisse2006}, allowing to consider a larger number of spins. Yet, for system sizes above $\approx30$, these brute-force approaches become unfeasible. Quantum Monte Carlo provides access to considerably larger systems \cite{Sandvik1991, Asselin2010}. However, it generates relevant statistical errors and becomes noticeably less accurate for frustrated systems. For frustrated systems, functional renormalization group for spin systems provide an alternative approach \cite{Reuther2014, Mueller2024}. Density matrix renormalization group (DMRG) is another powerful technique, especially when augmented by matrix product states and tensor network methods \cite{White1992,Daley2004, Orus2014, Cirac2021}. The method is able to compute finite-temperature observables by techniques such as matrix product density operators \cite{Verstraete2004, Zwolak2004}, purification \cite{Feiguin2005}, typical thermal states \cite{White2009, Stoudenmire2010}, thermal pure quantum states \cite{Sugiura2012, Iwaki2021} and exponential tensor renormalization group \cite{Chen2018}. It is most efficient for selected geometries -- spin chains and star-like lattices. Accessing higher-dimensional systems requires more computational effort \cite{Li2019, Gohlke2023}.
Hence, especially in 3D, alternative methods for simulating spin systems are of interest.

Recently, a new approach for computing dynamics in dense spin systems at infinite temperature was developed \cite{Graesser2021, Graesser2023}. It was termed dynamical mean-field theory for spins (spinDMFT) for its similarities to dynamical mean-field theory for fermions, which is widely used to study strongly-correlated electron systems \cite{Georges1996, Kotliar2004}. Both DMFTs use a single-site approximation and a self-consistency condition, and crucially utilize a time-dependent field. SpinDMFT has since been successfully applied to explain the results of NMR measurements, where the energy scales of the spins are small with respect to the thermal energy scale \cite{Graesser2024}. However for other applications, the infinite-temperature constraint is a clear limitation of the method.

In this work, we develop a novel method based on the method of Ref. \cite{Graesser2021}, which is applicable to spin ensembles at finite temperature. Similarly to the infinite-temperature approach, we use a time-dependent, Gaussian distributed mean field and a self-consistency connecting the correlations of the spins to the parameters of the mean-field probability distribution. However, we consider mean fields dependent on imaginary time, which turns out to be crucial for correctly modeling the thermal density operator. Furthermore, we propose a modified self-consistency condition including spin expectation values which can be non-zero in finite-temperature systems, for instance due to external fields or spontaneous symmetry breaking. We find that our method closely resembles a mean-field approach to the Sherrington-Kirkpatrick model \cite{Bray1980, Grempel1998, Georges2000}, despite an entirely different derivation.

We note that our approach relates to other mean-field theories in the literature. For example, a dynamical mean-field approach for spins (termed s-DMFT) has also been established in Ref.~\cite{Otsuki2013_2}. It is also based on a $1/z$ expansion and accesses finite temperatures and ordered phases. There are several conceptual differences between the two methods. Derivation-wise, the couplings in s-DMFT are scaled like $J\propto1/z$, in contrast to $J\propto 1/\sqrt{z}$ in our approach. Because of that, the approaches are tailored toward different systems. While spinDMFT is more suited to describe disordered spins, especially in systems with different signs of couplings, 
s-DMFT is designed for ordered systems. In Ref.~\cite{Otsuki2013_2}, the resulting impurity Hamiltonian introduces a bosonic field instead of a classical mean-field as in our case. This makes a solution of the impurity problem in s-DMFT more demanding numerically and computationally. On the other hand, s-DMFT computes correlations in momentum space, which is an advantage over the method we propose. We also highlight another recently developed mean-field approach termed density matrix mean-field theory (DMMFT) \cite{Zhang2024}. It is a general framework for constructing effective Hamiltonians, which has been successfully applied to spin systems as well. However, its focus is the description of phase transitions, while in this work our main focus is the dynamical behavior.

The structure of the article is as follows: In Sec. \ref{sec:derivation}, we derive spinDMFT for systems at finite temperature. We discuss the considered model and provide details of the single-site approximation and self-consistency. Then, in Sec. \ref{sec:comparison}, we provide a description of the numerical implementation of the method and benchmark it by comparison to results in finite-size systems. We also discuss the appearance of ferromagnetic order in spinDMFT. Finally, in Sec. \ref{sec:conclusion}, we conclude and give an overview of the next developments of finite-temperature spinDMFT. In the appendices, one can find the details of the derivation and implementation and a description of the numerical methods used for comparison.

\section{Derivation of the method}\label{sec:derivation}
\subsection{Model and definitions}
We consider an isotropic Heisenberg model for spins with $S=\frac{1}{2}$ in a static, homogeneous magnetic field
\begin{equation}\label{eq:full Hamiltonian}
    \mathbf{H} = \sum_{i<j}J_{ij}\vec{\mathbf{S}}_i\cdot\vec{\mathbf{S}}_j + \vec{B}\cdot\sum_i\vec{\mathbf{S}}_i.
\end{equation}
In this work, we denote operators acting on the Hilbert space by bold symbols and three-dimensional vectors by symbols with arrows above. The lattice geometry is described by the symmetric coupling matrix $J_{ij}$. Note that the geometry is completely general, in particular not restricted to periodic lattices. We introduce the moments of the coupling constants
\begin{equation}
    \mathcal{J}_{n,i} := \left(\sum_{j\neq i}|J_{ij}|^n\right)^{1/n}
\end{equation}
and two energy scales which will turn out to be relevant in the following considerations
\begin{align}
    \JL &:= \sum_{i\neq0}J_{0i}, & \JQ &:= \mathcal{J}_{2,0},
    \label{eq:couplingconstants}
\end{align}
which we term the linear and quadratic coupling constants. Here, we assume them to be site-independent as is the case for translation invariant
systems. Additionally, we define the effective coordination numbers for site $i$ as
\begin{align}
    z_i &:= \frac{\mathcal{J}_{1,i}^2}{\mathcal{J}_{2,i}^2}, & z_i' &:= \frac{\mathcal{J}_{2,i}^4}{\mathcal{J}_{4,i}^4}.
\end{align}
In the case of constant nearest-neighbor couplings, the effective coordination numbers coincide with the usual coordination number $z$ describing the number of neighbors of spin $i$, $z_i$ = $z_i'$ = $z$. Mean-field approaches become exact in the limit of an infinite coordination number. Accordingly, spinDMFT will be argued to be exact when $z_i,\, z_i'\rightarrow \infty$. This limit is realized in two cases: either the lattice becoming infinite-dimensional, or the interaction between spins being infinitely long-ranged. For the infinite-dimensional limit to exist, the couplings $J_{ij}$ need to scale as $J_{ij}\propto \nicefrac{1}{\sqrt{z}}$. This is plausible in the paramagnetic phase
\cite{Haule2002, Otsuki2013_1} and also used in the paradigmatic spin glass model, the Sherrington-Kirkpatrick model \cite{Bray1980, Grempel1998}.
The reason is that the relevant energy scale for a spin at site $i$ is $J_Q$, as will become clear in section \ref{sec:auxmodel}. To keep it constant in the $z\rightarrow\infty$ limit, the rescaling is required. 
Throughout this article we will assume that a finite $\JQ$ and $\JL$ do exist in the limit $z\rightarrow\infty$. This clearly requires that couplings with both signs, ferromagnetic and antiferromagnetic, must occur. By extension, the resulting approach will also be used as an approximation in finite-dimensional systems which have couplings only with one sign, but have finite values of $\JQ$ and $\JL$.
The dominant sign of the couplings, i.e., the sign of $\JL$, determines the behavior of the system as the temperature drops -- negative (positive) couplings favor spins (anti-)aligning, favoring (anti-)ferromagnetic ordering.

Our method determines imaginary-time spin dynamics for systems at finite temperature. We work in the thermal equilibrium so that the expectation values are computed as
\begin{equation}
    \left<\mathbf{X}\right> = \frac{1}{Z}\mathrm{Tr}\left(e^{-\beta\mathbf{H}}\mathbf{X}\right),
\end{equation}
with the inverse temperature $\beta={1/k_BT}$ and the partition function $Z = \mathrm{Tr}\,\mathrm{exp}(-\beta\mathbf{H})$. Imaginary-time evolution is carried out by the evolution operator $\mathbf{U}(\tau) = \mathrm{exp}(-\tau\mathbf{H})$ acting on operators as
\begin{equation}
    \mathbf{X}(\tau) = \mathbf{U}^{-1}(\tau)\mathbf{X}\mathbf{U}(\tau)
\end{equation}
with $\tau\in\mathbb{R}$. The main quantities of interest are the time-ordered imaginary-time spin correlations defined as
\begin{equation}\label{eq:correlations general}
    g_{ij}^{ab}(\tau_1,\tau_2) = \left<\mathcal{T}\mathbf{S}_i^a(\tau_1)\mathbf{S}_j^b(\tau_2)\right>,
\end{equation}
with $\tau_i \in[0,\beta]$. 
The indices $a,\,b$ represent the spin components $x,\, y$ and $z$. 
The time-ordering operator $\mathcal{T}$ orders operators such that the time increases from right to left.

As for infinite-temperature spinDMFT, it is useful to introduce the environment field of a spin $i$,
\begin{equation}\label{eq:environment field}
    \vec{\mathbf{V}}_i := \sum_{j\neq i} J_{ij}\vec{\mathbf{S}}_j.
\end{equation}
Furthermore, we split the Hamiltonian into a local part $\mathbf{H}_0$ consisting of all the terms containing $\vec{\mathbf{S}}_0$ and a rest term $\mathbf{W}$ according to
\begin{subequations}
\begin{align}
    \mathbf{H} &= \mathbf{H}_0 +\mathbf{W},\label{eq:split H}\\ 
    \mathbf{H}_0 &= \left(\vec{\mathbf{V}}_0+\vec{B}\right)\cdot\vec{\mathbf{S}}_0,\label{eq:H0}\\ 
    \mathbf{W} &= \sum_{ i,j\neq0,i<j}J_{ij}\vec{\mathbf{S}}_{i}\cdot\vec{\mathbf{S}}_j + \vec{B}\cdot\sum_{i\neq0}\vec{\mathbf{S}}_i.\label{eq:W}
\end{align}
\end{subequations}
The operator $\mathbf{W}$ corresponds to the Hamiltonian for a system with $\vec{\mathbf{S}}_0$ removed and will henceforth be referred to as the \enquote{cavity Hamiltonian}, cf.\ Ref.\ \cite{Georges1996}.

\subsection{Trotterization and effective time dependence}\label{sec:trotterization}

We are aiming to describe the dynamics of a single spin $\vec{\mathbf{S}}_0$. In particular, we want to determine the spin's expectation values as well as its autocorrelations defined as
\begin{subequations}
\label{eq:expval_autocorr_Z_definition}
\begin{align}
    \left<\mathbf{S}^{a}_{0}\right> &= \frac{1}{Z} \trace{ \oU(\beta) \mathbf{S}_0^a }, \\
    g^{ab}(\tau) := g^{ab}_{00}(\tau,0) &= \frac{1}{Z} \trace{ \mathcal{T}\oU(\beta)\oU^{-1}(\tau)\mathbf{S}_0^a \oU(\tau) \mathbf{S}_0^b }, \\
    Z &= \trace{ \oU(\beta) }.
\end{align}
\end{subequations}
Using Trotter's formula, the evolution operator can be written as
\begin{subequations}
\begin{align}
\label{eq:trotterization}
    \mathbf{U}(\tau) =& \lim_{n\rightarrow\infty}\prod_{k=1}^n e^{-(\tau/n)\mathbf{H}_0}e^{-(\tau/n)\mathbf{W}}\\ 
    =& e^{-\tau\mathbf{W}} \lim_{n\rightarrow\infty}\prod_{k=1}^n e^{(k\tau/n)\mathbf{W}}e^{-(\tau/n)\mathbf{H}_0}e^{-(k\tau/n)\mathbf{W}} \\
    =& e^{-\tau\mathbf{W}} \lim_{n\rightarrow\infty}\prod_{k=1}^n e^{-\frac{\tau}{n}\left(\vec{\mathbf{V}}_0(k\tau/n)+\vec{B}\right)\cdot \vec{\mathbf{S}}_0} \\
    =& e^{-\tau\mathbf{W}}\mathcal{T}e^{-\int_0^\tau\left(\vec{\mathbf{V}}_0(\tau')+\vec{B}\right)\cdot \vec{\mathbf{S}}_0 d\tau'},
\end{align}
\end{subequations}
where we introduced an effective time dependence of the environment fields with respect to the cavity Hamiltonian according to
\begin{equation}\label{eq:env fields evol}
    \vec{\mathbf{V}}_i(\tau) := e^{\tau\mathbf{W}}\vec{\mathbf{V}}_ie^{-\tau\mathbf{W}}.
\end{equation}
Note that the time ordering operator orders the terms $\vec{\mathbf{V}}(\tau)\cdot \vec{\mathbf{S}}_0$ with respect to the argument of $\vec{\mathbf{V}}$, with terms with larger $\tau$ appearing to the left. Inserting Eq.~\eqref{eq:trotterization} into Eq.~\eqref{eq:expval_autocorr_Z_definition} for all propagators we arrive at
\begin{subequations}\label{eq:expval_autocorr_Z_after_trotterization}
\begin{align}
    \left<\mathbf{S}^{a}_{0}\right> &= \frac1{Z} \trace{ e^{-\beta\mathbf{W}}\mathcal{T}e^{-\int^\beta_0\left(\vec{\mathbf{V}}_0(\tau')+\vec{B}\right)\cdot \vec{\mathbf{S}}_0 d\tau'} \mathbf{S}^{a}_{0}}, \\
    g^{ab}(\tau) &= \frac{1}{Z}\trace{ e^{-\beta\mathbf{W}}\mathcal{T}e^{-\int^\beta_\tau\left(\vec{\mathbf{V}}_0(\tau')+\vec{B}\right)\cdot \vec{\mathbf{S}}_0 d\tau'}\mathbf{S}^a_0e^{-\int_0^\tau\left(\vec{\mathbf{V}}_0(\tau')+\vec{B}\right)\cdot \vec{\mathbf{S}}_0 d\tau'}\mathbf{S}^b_0 }, \\
    Z &= \trace{ e^{-\beta\mathbf{W}}\mathcal{T}e^{-\int^\beta_0\left(\vec{\mathbf{V}}_0(\tau')+\vec{B}\right)\cdot \vec{\mathbf{S}}_0 d\tau'} },
\end{align}
\end{subequations}
where we employed that $e^{-\tau\mathbf{W}}$ commutes with $\mathbf{S}_0^a$ and combined the first two propagators in the expression for the autocorrelation. Here, the advantage of using imaginary time evolution becomes clear. The thermal density operator and the evolution operator are related by $e^{-\beta\mathbf{H}}=\mathbf{U}(\beta)$ allowing us to consider an evolution with respect to a real-valued parameter $\tau'$ only. In case of real-time correlations, combining time evolution and density operator would require us to introduce a time evolution with in complex time along a Keldysh contour implying a more involved algorithm.

Note that up to this point no approximation has been made. Eqs.~\eqref{eq:expval_autocorr_Z_after_trotterization} still form a quantum many-body problem which is not exactly solvable. However, from this formulation the structure of the effective single-site model defined by spinDMFT can already be guessed. 
The time evolution of spin $\vec{\mathbf{S}}_0$ is generated by its coupling to the magnetic field and the environment field. 
The latter obtains an effective time dependence due to the remaining lattice through the operator $\oW$ and will be substituted by the mean field.

\subsection{Mean-field approximation}\label{sec:auxmodel}
We aim to replace the environment fields in Eq. \eqref{eq:expval_autocorr_Z_after_trotterization} by a classical, time dependent mean-field. The argument is as follows: as $z\rightarrow\infty$, the couplings approach zero, making the influence of each neighbor smaller. In this limit, the environment field consists of a large number of small, independent contributions (for a more detailed line of argument, see appendix \ref{ap:cavity_method}). 
Therefore, we are justified in replacing the full many-body problem by an auxiliary model
\begin{equation}
    \mathbf{H}_{\mathrm{mf}}(\tau) =\left(\vec{V}(\tau)+\vec{B}\right)\cdot\vec{\mathbf{S}}_0
\end{equation}
with a Gaussian distributed classical mean field $\vec{V}(\tau)$. 
In this framework, expectation values are computed by performing a quantum average in the single-site model and a classical average over the probability space of the mean fields. This entails
\begin{subequations}
\label{eq:aux_model_traces}
\begin{align}
    \left<\mathbf{S}^{a}_{0}\right> &= \frac{1}{Z_0}\int\mathcal{D}\vec{\mathcal{V}}\,p[\vec{\mathcal{V}}]\trace{ \mathcal{T}e^{-\int^\beta_0\mathbf{H}_{\mathrm{mf}}(\tau')d\tau'} \mathbf{S}^{a}_{0} }, \\
    g^{ab}(\tau) &= \frac{1}{Z_0}\int\mathcal{D}\vec{\mathcal{V}}\,p[\vec{\mathcal{V}}]\trace{ \mathcal{T}e^{-\int^\beta_\tau\mathbf{H}_{\mathrm{mf}}(\tau')d\tau'}\mathbf{S}^a_0e^{-\int_0^\tau \mathbf{H}_{\mathrm{mf}}(\tau')d\tau'}\mathbf{S}^b_0 }, \label{eq:spin_corr_mf} \\
    Z_0 &= \int\mathcal{D}\vec{\mathcal{V}}\,p[\vec{\mathcal{V}}]\trace{ \mathcal{T}e^{-\int^\beta_0\mathbf{H}_{\mathrm{mf}}(\tau')d\tau'} }. \label{eq:partition_mf}
\end{align}
\end{subequations}
These expressions are equivalent to the ones of the original lattice problem if the classical moments of the mean field match the quantum expectation values of the original field $\mathbf{V}_0$. This is expressed in the following conditions
 \begin{subequations}
\begin{align}
    \overline{V^a(\tau_1)} &= \left<\mathbf{V}_0^a(\tau_1)\right>,
    \label{eq:mfaverages_raw} \\
    \mathrm{Cov}\left\{V^a(\tau_1),V^b(\tau_2)\right\} &= \mathrm{Re}\left<\mathcal{T}\mathbf{V}_0^a(\tau_1)\mathbf{V}_0^b(\tau_2)\right> - \left<\mathbf{V}_0^a(\tau_1)\right>\left<\mathbf{V}_0^b(\tau_2)\right>.
    \label{eq:mfcovariances_raw}
\end{align}
\end{subequations}
The real part in Eq. \eqref{eq:mfcovariances_raw} is taken to ensure the covariance matrix is symmetric and the mean-fields are real. The imaginary part reflects the non-commutativity of the involved operators. But the environment operators behave classically in the limit $z\rightarrow\infty$. Hence, we assume that the imaginary part does not play a role for the distribution of the mean fields. Inserting the definition of the environment fields \eqref{eq:environment field}, we obtain
\begin{subequations}
\label{eq:mfselfcons}
\begin{align}
    \overline{V^a(\tau_1)} &= \sum_{i\neq0} J_{0i} \left<\mathbf{S}_i^a(\tau_1)\right> = \JL \left<\mathbf{S}_0^a(\tau_1)\right>,
    \label{eq:mfaverages} \\
    \mathrm{Cov}\left\{V^a(\tau_1),V^b(\tau_2)\right\} &= \sum_{i,j\neq 0}J_{0i}J_{0j}\left(\mathrm{Re}\left<\mathcal{T}\mathbf{S}_i^a(\tau_1)\mathbf{S}_j^b(\tau_2)\right> - \left<\mathbf{S}_i^a(\tau_1)\right>\left<\mathbf{S}_j^b(\tau_2)\right>\right)\notag\\
    &\approx \sum_{i\neq 0}J_{0i}^2\left(\mathrm{Re}\,g_{ii}^{ab}(\tau_1-\tau_2) - \left<\mathbf{S}_i^a\right>\left<\mathbf{S}_i^b\right>\right)\notag\\
    &= \JQ^2\left(\mathrm{Re}\,g^{ab}(\tau_1-\tau_2)-\left<\mathbf{S}_0^a\right>\left<\mathbf{S}_0^b\right>\right).
    \label{eq:mfcovariances}
\end{align}
\end{subequations}
In the second step of Eq.~\eqref{eq:mfaverages} and the third step of Eq.~\eqref{eq:mfcovariances}, we assumed homogeneity of the spin ensemble 
and used the linear and quadratic coupling constants defined in Eq.~\eqref{eq:couplingconstants}. 
The assumption of homogeneity may not be justified if the translational symmetry is spontaneously broken, as in antiferromagnets on bipartite lattices.

In the second step of  Eq.~\eqref{eq:mfcovariances}, we discarded contributions with $i\neq j$. The validity of this single-site approximation has been proven in the case of infinite temperature \cite{Graesser2021} 
and in the case of the Sherrington-Kirkpatrick model \cite{Bray1980} and we use it by extension in our present approach. The connection of our method to spin glass physics will be further explored in section \ref{sec:comparison numerics} and appendix \ref{ap:grempel}. The linear sum of the couplings can take any real value, leading to a ferromagnetic (antiferromagnetic) system when the negative (positive) couplings dominate, meaning $J_L<0$ ($J_L>0$). 
We will see that the approach provides a surprisingly good approximation even for the dynamics of a system with a finite number of same-sign couplings
when discussing the results in section \ref{sec:comparison}.


Equation~\eqref{eq:mfselfcons} represents a self-consistency condition because the spin autocorrelations and expectation values on the right-hand-side are unknown. In the single-site model, they can be calculated through Eq.~\eqref{eq:aux_model_traces} requiring the mean-field distribution functional and, thus, the moments on the left-hand-side of Eq.~\eqref{eq:mfcovariances}. This self-consistency problem can be solved by numerical iteration: Given an initial guess for the spin autocorrelations and expectation values, the moments of the mean-field distribution are obtained. With the distribution given, new spin autocorrelations and expectation values are computed using Eq.~\eqref{eq:aux_model_traces}. These are then reinserted to Eq.~\eqref{eq:mfcovariances} to obtain new mean-field moments and the procedure is iterated until convergence is achieved, i.e., until the difference between the correlations in two subsequent steps is smaller than a certain tolerance value. Details on the numerical implementation will be provided in the next section.

The derivation and the resulting single-site model is similar to infinite-temperature spinDMFT \cite{Graesser2021}, but we emphasize a few crucial differences: First, we work in imaginary time on a finite interval $\tau\in[0,\beta)$. Second, we explicitly take the real part of the environment field correlations in Eq. \eqref{eq:mfcovariances_raw} to ensure that the time ordering is symmetric and the mean fields are real. Third, we need to take finite spin expectation values into account, which can arise from spontaneous symmetry breaking and/or from external magnetic fields. These expectation values affect the mean-field averages and thereby also the spin dynamics.

\section{Implementation and benchmarks}
\label{sec:comparison}
\subsection{Numerical implementation}

In order to test the method, we use a numerical procedure to solve the self-consistency condition outlined in the section above, utilizing a similar approach to the implementation of infinite-temperature spinDMFT \cite{Graesser2021}. The main challenge is computing the path integral over mean-field configurations for spin correlations \eqref{eq:spin_corr_mf}. To this end, we discretize imaginary time $\tau\in[0,\beta]$ 
into equidistant steps $(\tau_0=0, \tau_1,\ldots\tau_L=\beta)$ with $\tau_l=l\frac{\beta}{L}$. 
Then, the mean-field configuration can be written as a $3(L+1)$-dimensional vector $\vec{\mathcal{V}}=(\vec{V}(\tau_0),\vec{V}(\tau_1),\ldots,\vec{V}(\tau_L))^T$ and the path integral becomes a $3(L+1)$-dimensional integral
\begin{equation}
    \int\mathcal{D}\vec{\mathcal{V}} = \int d^3\vec{V}(\tau_0)\int d^3\vec{V}(\tau_1)\ldots\int d^3\vec{V}(\tau_L).
\end{equation}
The probability distribution of the mean fields takes the form
\begin{equation}\label{eq:prob distr}
    p[\vec{\mathcal{V}}]=\frac{1}{\sqrt{\mathrm{det}2\pi\mathcal{M}}}\mathrm{exp}\left[-\frac{1}{2}\left(\vec{\mathcal{V}}-\vec{\mathcal{U}}\right)^T\mathcal{M}^{-1}\left(\vec{\mathcal{V}}-\vec{\mathcal{U}}\right)\right],
\end{equation}
where the covariance matrix is defined as
\begin{equation}\label{eq:cov matrix}
    \mathcal{M}_{\tau_1\tau_2}^{ab} = \mathrm{Cov}\left\{V^a(\tau_1),V^b(\tau_2)\right\}
\end{equation}
and the average vector is
\begin{equation}\label{eq:average vector}
    \mathcal{U}^a_{\tau}=\overline{V^a(\tau)}.
\end{equation}
Note that the average vector is constant in time due to time translational invariance. The covariance matrix is $3(L+1)$-dimensional. To evaluate the high-dimensional integrals required to compute spin correlations and expectation values, we use the Monte Carlo approach. We draw mean-field configurations from the probability distribution $p[\vec{\mathcal{V}}]$, calculate the resulting expectation values of the observables of interest, and take the arithmetic average of many single-sample results to obtain the final values. The number of samples is taken sufficiently large to keep the statistical error below the desired threshold. For details of the implementation, i.e., how the sampling and time evolution is performed numerically, see appendix \ref{ap:spindmft implementation}.

\subsection{Spin correlations}\label{sec:spin correlations}
Before proceeding to the results of the algorithm, we analyze the mathematical structure of the computed quantities. The properties of spin operators imply that
\begin{equation}
    g^{ab}(0) = \frac{1}{4}\delta^{ab} + \frac{\mathrm{i}}{2}\varepsilon^{abc}\left<\mathbf{S}_0^c\right>,
\end{equation}
where $\varepsilon^{abc}$ is the Levi-Civita symbol and the sum convention is implied. 
The diagonal correlations start at $\frac{1}{4}$, and the off-diagonal ones depend on the expectation value of the spin in the direction perpendicular to $a, b$. 
By taking the complex conjugate of the correlations, one obtains the following formulas
\begin{subequations}
\begin{align}
    \mathrm{Re}\,g^{ab}(\tau) &= \frac{1}{2}\left(g^{ab}(\tau) + g^{ba}(\tau)\right);\\
    \mathrm{Im}\,g^{ab}(\tau) &= \frac{1}{2\mathrm{i}}\left(g^{ab}(\tau) - g^{ba}(\tau)\right);\\
    g^{ab}(\tau) &= g^{ba}(\beta-\tau)\label{eq:beta symm}.
\end{align}
\end{subequations}
Note that for the diagonal correlations $g^{aa}(\tau)$, the imaginary parts vanish, and equation \eqref{eq:beta symm} implies that they are symmetric around $\tau=\nicefrac{\beta}{2}$: $g^{aa}(\beta-\tau)=g^{aa}(\tau)$. The presence of symmetry further simplifies the correlations. When $\vec{B}=0$, the system becomes invariant under spin rotations. Therefore, non-diagonal correlations vanish, $g^{ab}=0$ for $a\neq b$, and the diagonal correlations are equal $g^{xx}(\tau) = g^{yy}(\tau)=g^{zz}(\tau)$. Additionally, any spin expectation value vanishes as well, i.e., $\left<\vec{\mathbf{S}}_0^a\right>=0$. Therefore, it is sufficient to compute a single correlation $g^{xx}(\tau)$, which is real. In the presence of an external magnetic field $\vec{B}$ pointing along the $z$ axis, the rotational symmetry is reduced to the symmetry under rotations around the $z$ axis. Under the $\frac{\pi}{2}$-rotation around the $z$ axis, the spin components transform as $\mathbf{S}^x\rightarrow\mathbf{S}^y$ and $\mathbf{S}^y\rightarrow-\mathbf{S}^x$. Therefore, $g^{az}(\tau)=0$ for $a=x,y$, $g^{xx}(\tau)=g^{yy}(\tau)$ and $g^{xy}(\tau)=-g^{yx}(\tau)$. Then, there are three independent correlations: $g^{xx}(\tau)$, $g^{xy}(\tau)$ and $g^{zz}(\tau)$. Due to relation \eqref{eq:beta symm}, the $xy$ correlation is $\beta$-antisymmetric $g^{xy}(\beta-\tau) = -g^{xy}(\tau)$ and therefore purely imaginary.

\subsection{Comparison to finite-size results}\label{sec:comparison numerics}

To benchmark spinDMFT at finite temperatures, 
we compare the resulting spin correlations to those computed in finite-size spin systems. The finite-size approaches clearly cannot access the infinite-dimensional limit. The larger the lattice dimension, the more pronounced the resulting finite-size effects are. Still, these approaches represent the best benchmarks at hand and it will turn out that in certain cases spinDMFT yields remarkably good agreement.

We consider three system for comparison: a nearest-neighbor ferromagnet, a nearest-neighbor antiferromagnet and an infinite-range system with random couplings. For each system, we consider the isotropic case, i.e, without magnetic field, and the case with magnetic field. For the models with nearest-neighbor interactions, we consider a simple cubic lattice without frustration. For these systems, quantum Monte Carlo (QMC) methods are sign-problem-free and very efficient. We use the Discrete Space Quantum Systems Solver implementation of the directed loop algorithm (DSQSS-DLA \cite{Motoyama2021}) to compute imaginary-time correlations in a $32\times32\times32$ system with periodic boundary conditions. At the considered temperatures, we verified that for this system size the finite-size effects were smaller than the sampling errors. For the random-coupling system, we draw couplings between each pair of spins in the cluster from a Gaussian distribution centered on zero, and average over correlations from many coupling configurations. In each sample, we normalize the couplings so that $J_Q$ remains constant. Since the system is highly frustrated, QMC is not viable. Instead, we use the Chebyshev expansion technique (CET). It allows us to approximate the evolution operator by expanding it in terms of Chebyshev polynomials of the Hamiltonian \cite{Tal-Ezer1984, Kosloff1994, Weisse2006}. For a detailed description, see appendix \ref{ap:numerical methods}. We compute correlations for systems of $N=9,10,11,12,13$ and perform a $1/N$ extrapolation. For a discussion of the finite size effects, see appendix \ref{ap:real time}.

\begin{figure}
    \centering
    \includegraphics[width=\linewidth]{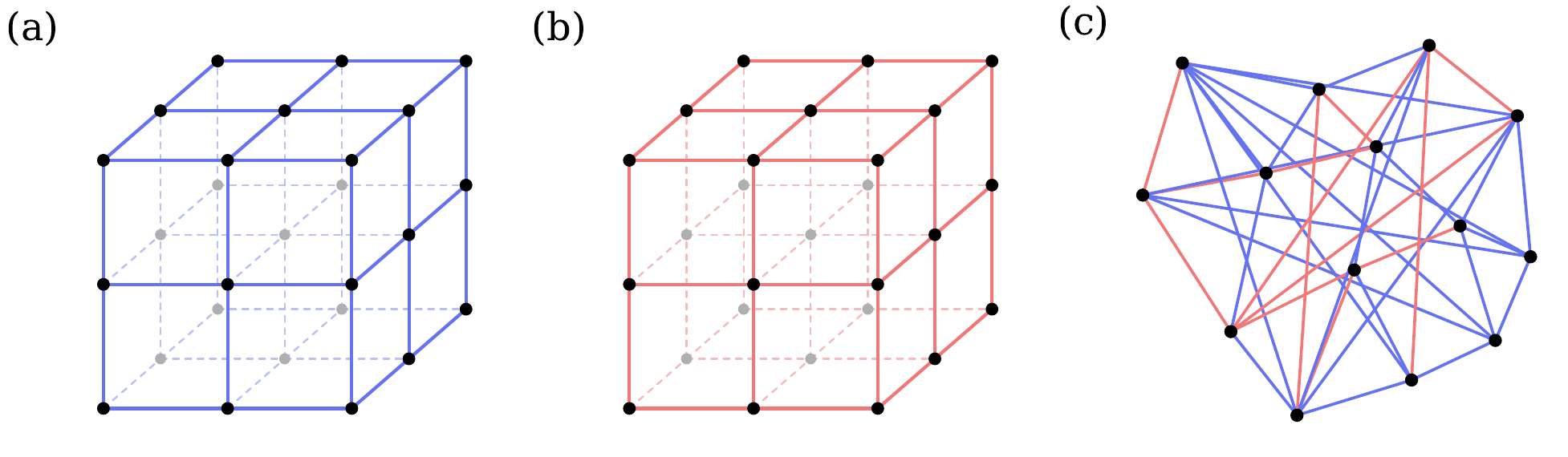}
    \caption{Three types of systems used for benchmarking: nearest-neighbor ferromagnet (a) and antiferromagnet (b) on a cubic lattice with periodic boundary conditions and (c) infinite-range system with random couplings. Black dots represent lattice sites and the blue (red) lines negative (positive) couplings which are ferromagnetic (antiferromagnetic) in character. Note that the system sizes are not representative of the ones used in the benchmarks.}
    \label{fig:lattices}
\end{figure}

\begin{figure}[H]
    \centering
    \includegraphics[width=\textwidth]{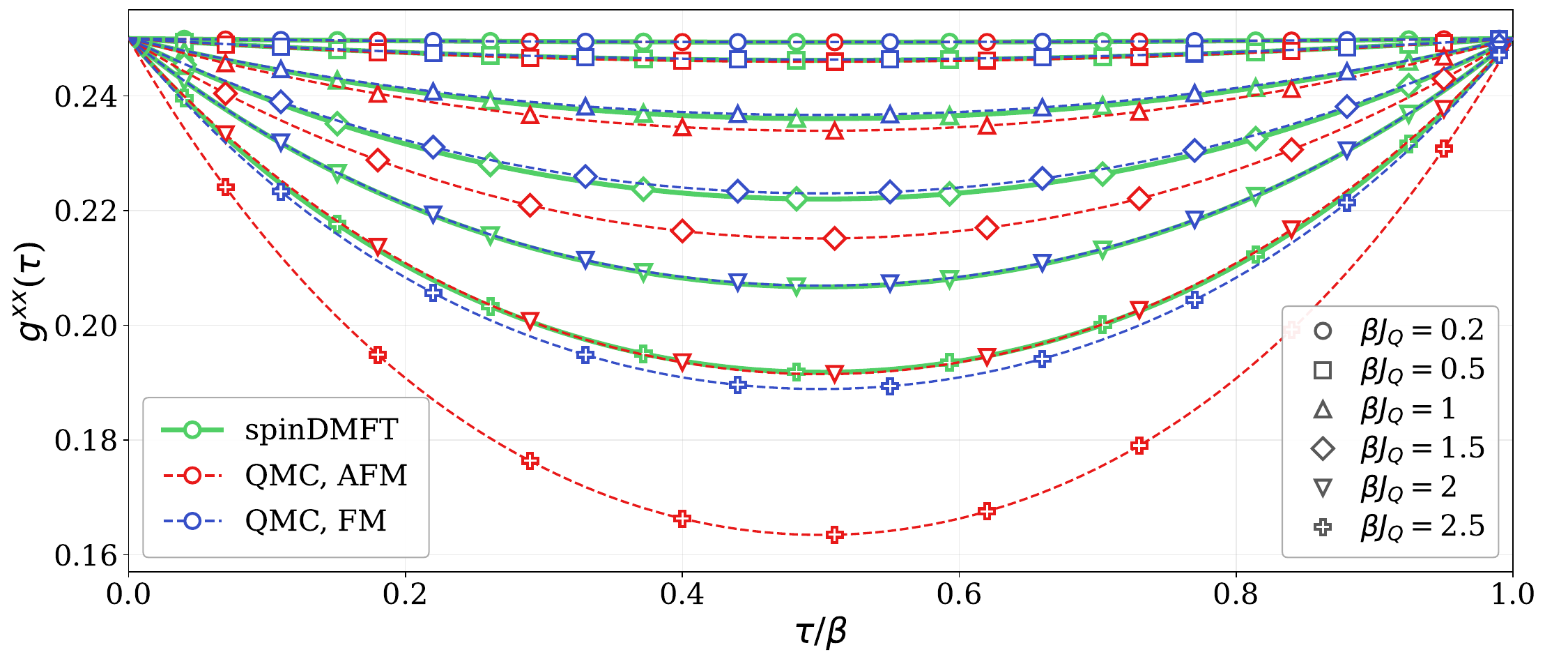}
    \caption{Comparison of correlations obtained in spinDMFT (green) and in a ferromagnet (blue) and antiferromagnet (red) at different temperatures, which are depicted by different symbols. The considered finite-size systems are a 3D cubic $32\times32\times32$ lattice with nearest-neighbor negative/positive couplings, see Fig. \ref{fig:lattices}. Note that there is an incidental overlap of the spinDMFT curve at $\beta J_Q=2.5$ and the AFM curve at $\beta J_Q = 2$. The error bars for spinDMFT and finite-size data are smaller than the width of the curves.}
    \label{fig:spindmft vs AFM FM}
\end{figure}

\begin{figure}[H]
    \centering
    \includegraphics[width=\textwidth]{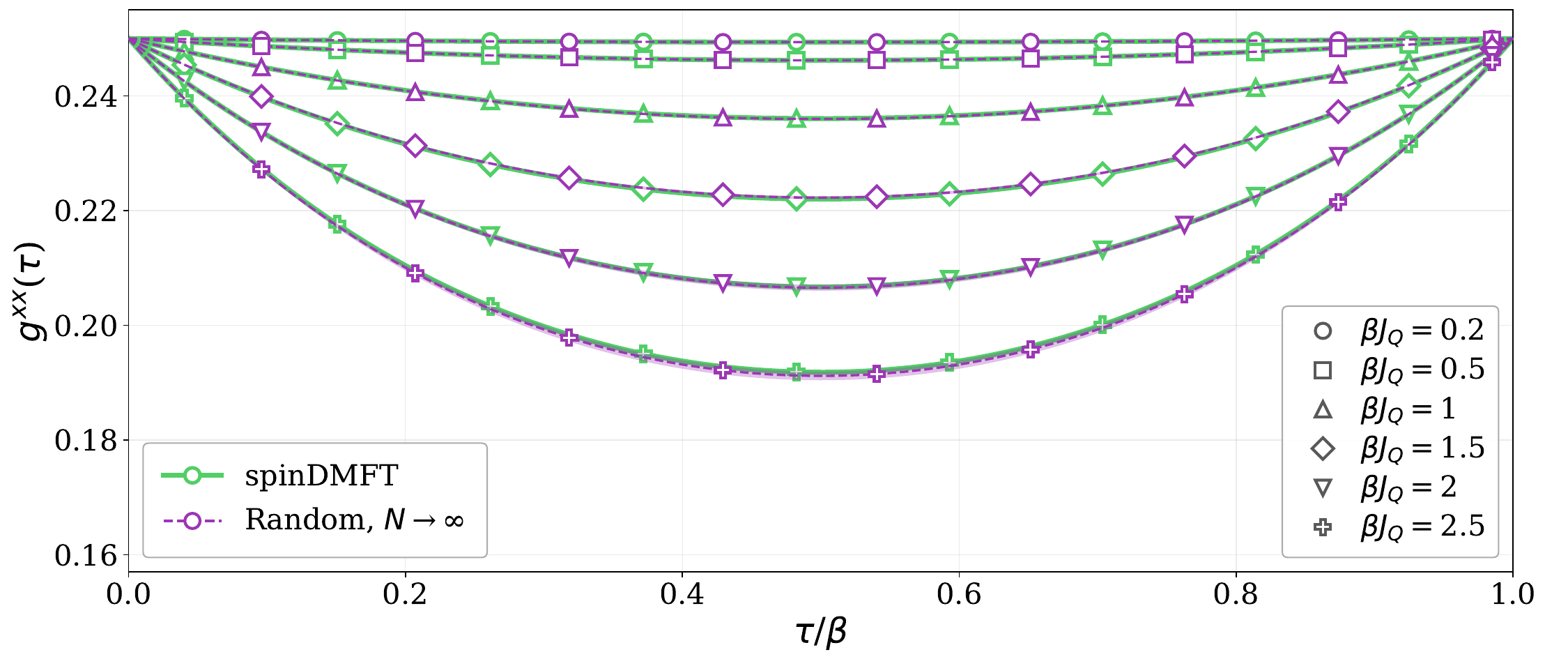}
    \caption{Comparison of correlations obtained in spinDMFT (green) and in an infinite-range system with random couplings (purple) at different temperatures (depicted by different symbols). The random-coupling data was obtained by a $1/N$ extrapolation from systems of size $N\in\{9,10,11,12,13\}$. The spinDMFT error bars are smaller than the width of the curves. In the random system there are small but noticeable deviations at the lowest temperature resulting from the extrapolation, i.e., from finite-size effects.}
    \label{fig:spindmft vs random}
\end{figure}

In Figs. \ref{fig:spindmft vs AFM FM} and \ref{fig:spindmft vs random}, we compare the correlations resulting from spinDMFT with the ones computed in finite-size systems at various temperatures. Since for $\vec{B}=0$ spinDMFT depends on the system only through the parameter $J_Q$, it cannot distinguish between different signs of couplings. Therefore, a single curve can be compared to all three finite-size systems. At very high temperatures, the agreement is very good in all cases, because then the signs of the couplings do not play a role and the single-site approximation is well-justified. As the temperature is decreased, the curves from the antiferromagnetic system start deviating considerably from the ones obtained by spinDMFT.
We attribute this behavior tentatively to incipient antiferromagnetic order and its spin flop transition
which are not captured by an approach that does not allow for breaking translational symmetry. In view of the strong deviations in the antiferromagnetic case, the correlations in the ferromagnetic system are surprisingly close to the ones from spinDMFT. Yet, also here deviations start to become visible as the temperature is lowered, see the results for $\beta J_Q \geq 0.5$. We conclude, that in its current formulation single-site spinDMFT is unable to describe the antiferromagnetic system, but captures the ferromagnet with nearest-neighbor interactions qualitatively for not too low
temperatures.

We observe that the curves obtained in spinDMFT always lie between the correlations in the ferromagnetic and antiferromagnetic system. This agrees with our expectations, since spinDMFT computes the correlations for a mixed-sign system. Therefore, including more positive/negative couplings should shift the correlations in opposite directions. For the random-coupling system, the correlations agree with spinDMFT very well. The deviations at the lowest considered temperature are likely due to an insufficient number of coupling configurations included in the averaging rather than from a shortcoming of spinDMFT. Therefore, we see that the method is able to precisely calculate correlations for a spin glass in the paramagnetic phase, thus reproducing the results of Ref. \cite{Grempel1998}. For a direct comparison of this, see appendix \ref{ap:grempel}.


\begin{figure}[H]
    \centering
    \includegraphics[width=\linewidth]{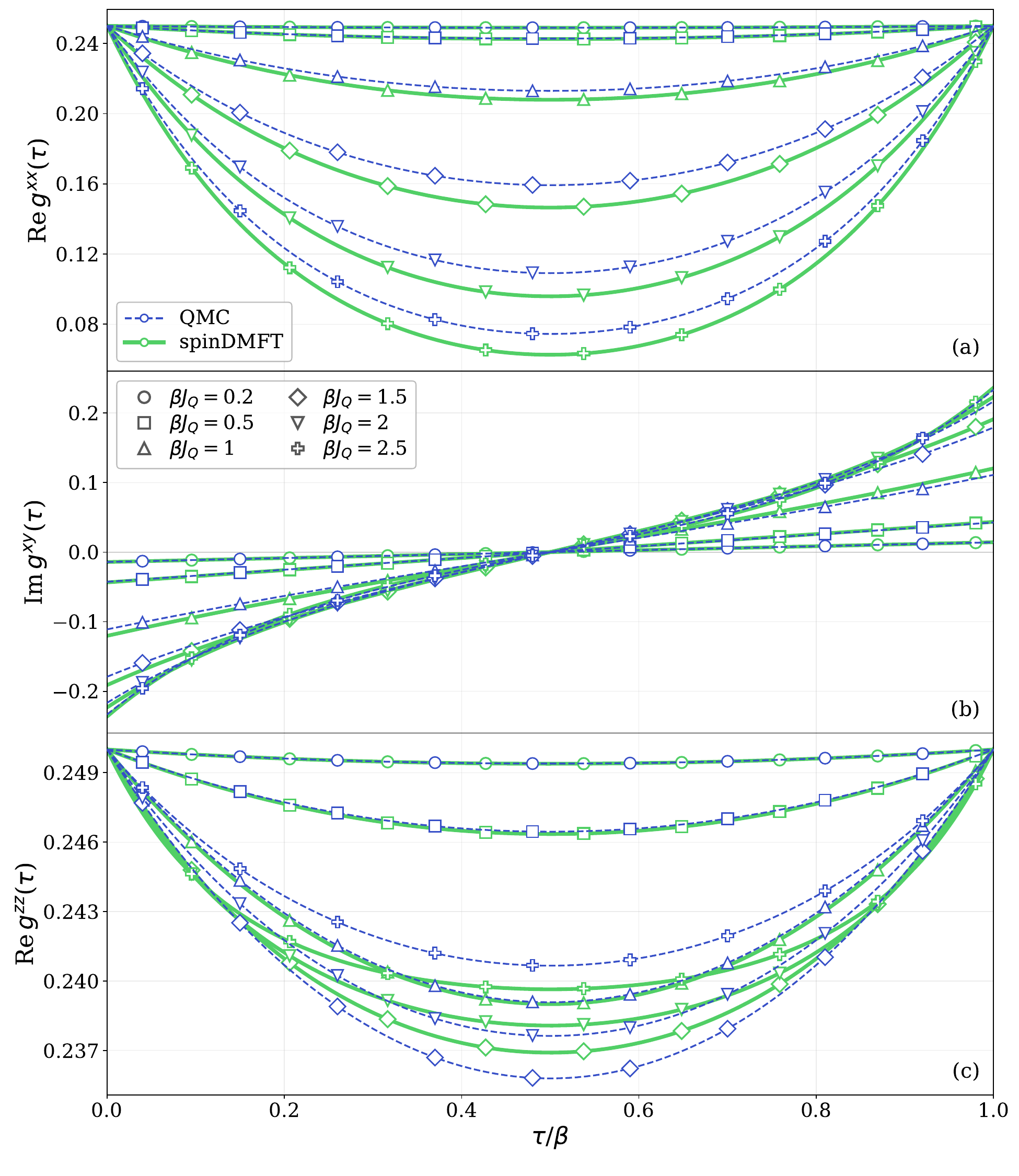}
    \caption{Comparison of correlations obtained in spinDMFT (green) and in a ferromagnetic system (blue) at different temperatures (signified by the symbols) in the presence of an external magnetic field ($B_z/J_Q=0.5$). We depict the $xx$ correlation in (a), $xy$ correlation in (b) and $zz$ correlation in (c). Note the very different $y$-axis scales used for (a) and (c). The considered finite-size system is a 3D cubic $32\times32\times32$ lattice with nearest-neighbor negative couplings. In (c), note the incidental overlap of the spinDMFT curves at $\beta J_Q=1.5,\,2$ and the FM curve at $\beta J_Q=2.5$. The error bars resulting from time discretization and averaging over the distribution of mean
    fields for spinDMFT are smaller than the line width of the curves.}
    \label{fig:FM B}
\end{figure}

In Fig. \ref{fig:FM B} we present a comparison between spinDMFT and the ferromagnetic system in a magnetic field applied in the $z$ direction. The linear coupling constant in spinDMFT was set to $J_L=-\sqrt{6}J_Q$, corresponding to $z=6$ neighbors with equal negative couplings. We see good agreement at high temperatures and increasing deviations as the temperature decreases. However, we note that the qualitative behavior of the correlations in spinDMFT and in the finite-size system is similar -- the $xx$ correlations take much smaller values than the $zz$ correlations, reflecting the tendency of spins to align with the magnetic field. Note that the values of the $zz$ correlations first decrease with temperature, and then start increasing for both spinDMFT and the exact results. This is expected, since for a ferromagnetic system at zero temperature the $zz$ correlation would stay constant at $\frac{1}{4}$. The $xy$ correlations take values close to the maximum value $\frac{1}{4}$, corresponding to a large magnetic moment developing even at low fields.

\begin{figure}[H]
    \centering
    \includegraphics[width=\linewidth]{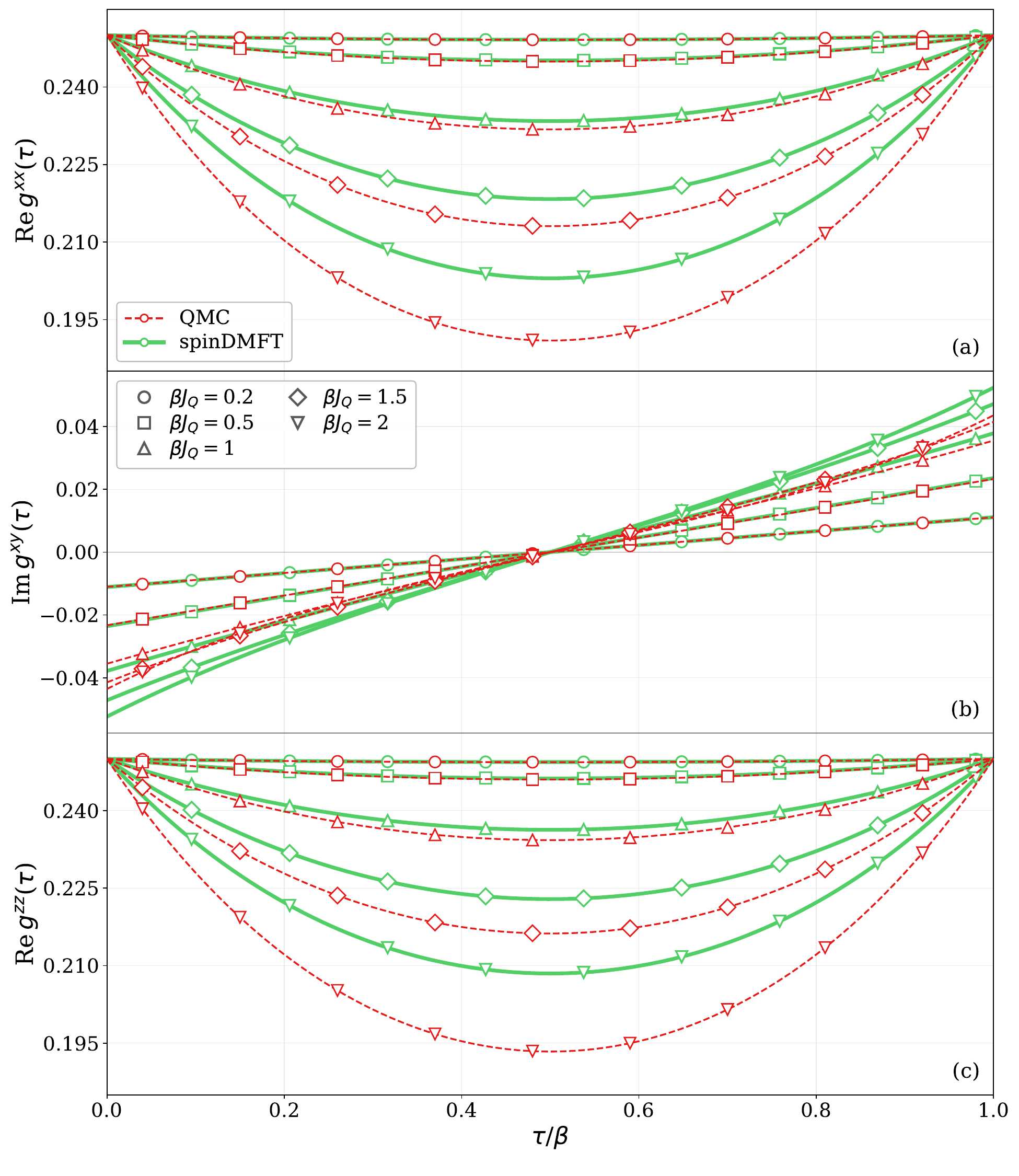}
    \caption{Comparison of correlations obtained in spinDMFT (green) and in an antiferromagnetic system (red) at different temperatures (signified by the symbols) in the presence of an external magnetic field ($ B_z/J_Q=0.5$). We depict the $xx$ correlation in (a), $xy$ correlation in (b) and $zz$ correlation in (c). The considered finite-size system is a 3D cubic $32\times32\times32$ lattice with nearest-neighbor positive couplings. The errorbars for spinDMFT and the finite-size data are smaller than the width of the curves.}
    \label{fig:AFM B}
\end{figure}

In Fig. \ref{fig:AFM B} we present a comparison between spinDMFT and the antiferromagnetic system in a magnetic field applied in the $z$ direction. The linear coupling constant in spinDMFT was set to $J_L=\sqrt{6}J_Q$, corresponding to $z=6$ neighbors with equal positive couplings. Once again, we see good agreement at high temperatures and increasing deviations as the temperature decreases. The qualitative behavior is as expected -- in both spinDMFT and exact results, $xx$ and $zz$ correlations have a similar magnitude, corresponding to the system resisting the influence of the magnetic field. The $xy$ correlations take small values for the same reason. In both cases, we see that spinDMFT is not quantitatively reliable, but can at least provide qualitative insight into the behavior of correlations in the small-$z$ systems.
At low temperatures $\beta J_Q > 2$ in the antiferromagnetic case, we found that spinDMFT does not converge, with the differences between consecutive results remaining large even after 20 iterations. We suspect it to be an indication of the insufficiency of the single-site approximation and the self-consistency condition \eqref{eq:mfaverages} to describe antiferromagnetic ordering. It is clear, that the approximation of the spin expectation values being equal on all sites is not valid for a system in which spins anti-align or cannot do so due to a spin flop transition. 
Since the method does not consider neighboring spins explicitly, it is not able to determine the parallel or antiparallel orientation of adjacent spins. However, this problem only arises when we include the external magnetic field, as only in this case he spins acquire expectation values due to the breaking of rotational symmetry. This in turn causes the sublattice symmetry to be spontaneously broken.

\begin{figure}[H]
    \centering
    \includegraphics[width=\linewidth]{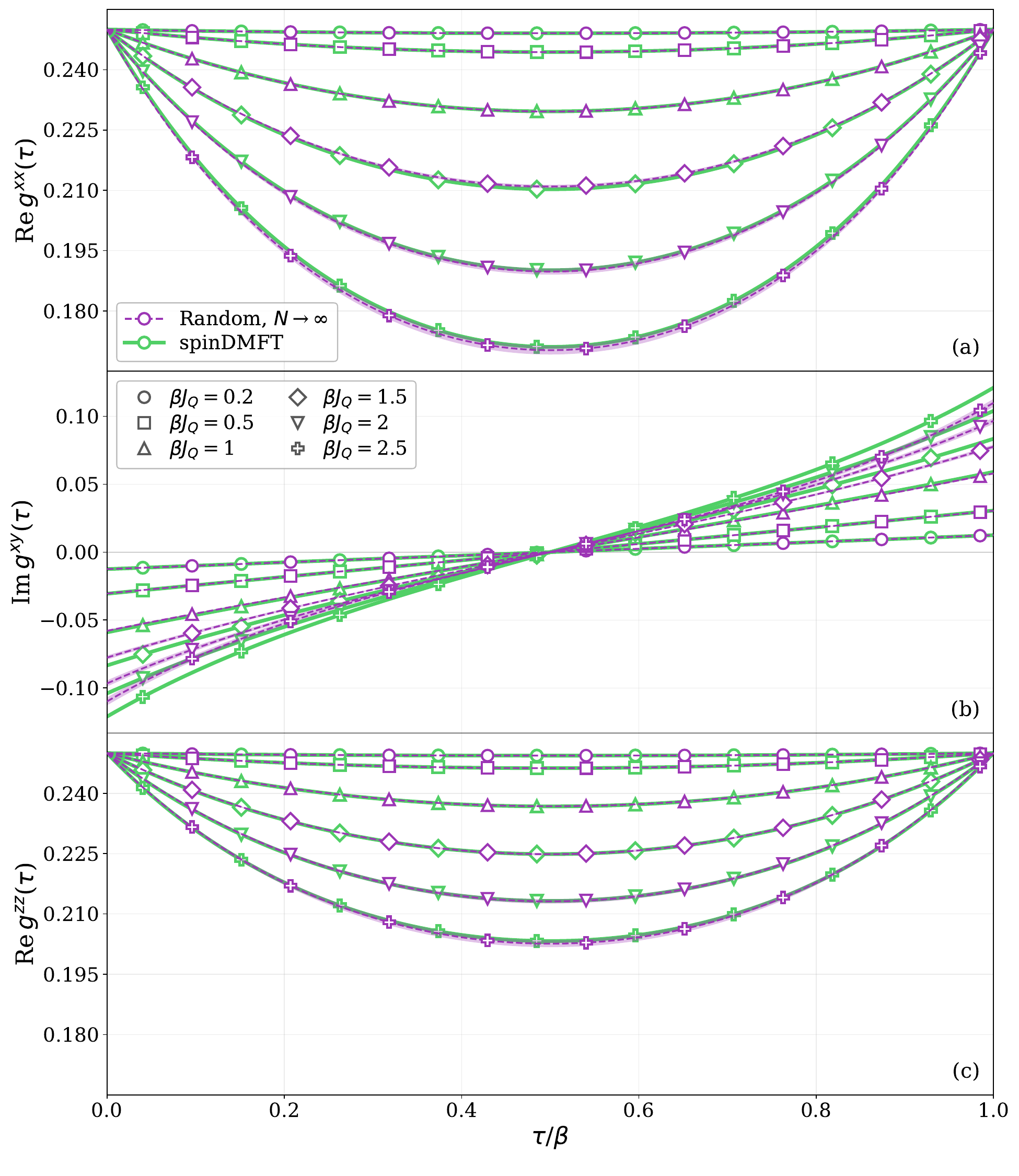}
    \caption{Comparison of correlations obtained in spinDMFT (green) and in an infinite-range system with random couplings (purple) at different temperatures (signified by the symbols) in the presence of an external magnetic field ($ B_z/J_Q=0.5$). We depict the $xx$ correlation in (a), $xy$ correlation in (b) and $zz$ correlation in (c). The random-coupling data was obtained by a $1/N$ extrapolation from systems of size $N\in\{9,10,11,12,13\}$. The spinDMFT error bars are smaller than the width of the curves. The random-coupling data has small, but discernible errors due to the extrapolation.}
    \label{fig:Random B}
\end{figure}

In Fig. \ref{fig:Random B} we present a comparison between spinDMFT and the random-coupling system in a magnetic field applied in the $z$ direction. The linear coupling constant in spinDMFT was set to $J_L=0$, since in the random-coupling system positive and negative couplings are equally probable so that they average to zero. As in the case without an external field, the spinDMFT describes the random-coupling system very well. The only noticeable deviations appear in the $xy$ correlation at low temperatures. This might indicate an insufficiency of the considered self-consistency condition, for instance by taking the real part of the correlations. In the current form, the $xy$ correlations do not enter into the self-consistency at all and may not be simulated correctly. This question deserves further studies, but this is beyond the scope of the present article.

\subsection{Ferromagnetic phase transition in spinDMFT}\label{sec:FM phase transition}

\begin{figure}[H]
    \centering
    \includegraphics[width=0.75\textwidth]{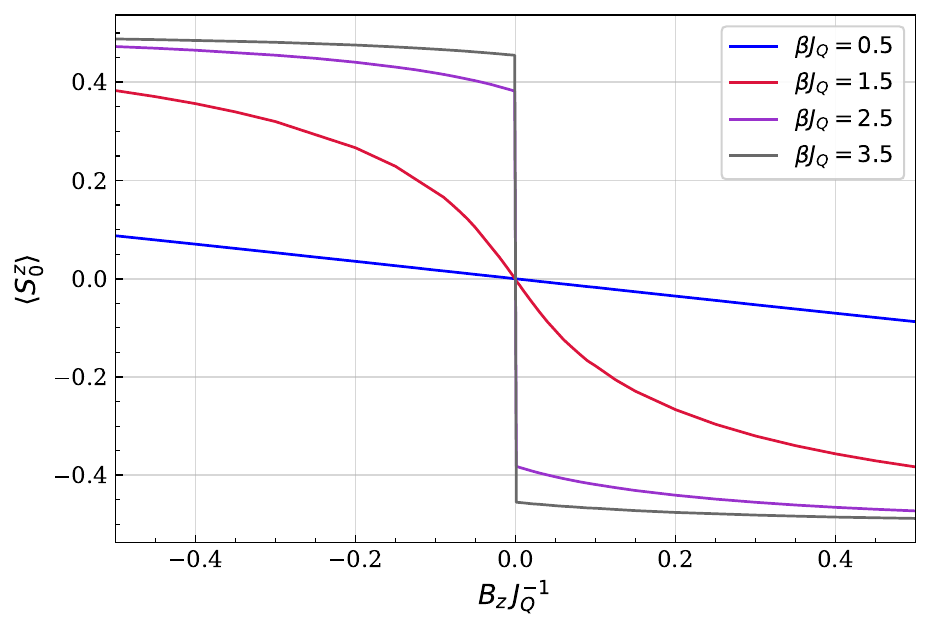}
    \caption{Dependence of the spin expectation value on the magnetic field in ferromagnetic spinDMFT at a range of temperatures.}
    \label{fig:M of B}
\end{figure}

In this section, we want to study the dependence of the spin expectation values on the external magnetic field and the possibility of order arising in spinDMFT. We take $J_L=-\sqrt{6}J_Q$ to model the ferromagnetic system on a cubic lattice. At each temperature, we run the simulation using the following procedure: first, we generate results for a large magnetic field $B_z=0.5J_Q$. Then, we use the obtained data as an initial condition for smaller magnetic field. The smallest considered field was $h_z=0.001J_Q$; we keep a non-zero field 
to suppress spurious numerical oscillation of the magnetization. In Fig. \ref{fig:M of B}, the obtained magnetization curves at three different temperatures are shown. One can see, that for the two higher temperatures, $M\rightarrow0$ as $B\rightarrow0$. However, as the temperature is lowered, $\langle\mathbf{S}^z_0\rangle$ remains nonzero as the magnetic field vanishes. We see a clear hallmark of a ferromagnetic phase transition.
This provides evidence that we can capture the ferromagnetic ordering transition for $\JL<0$. It is contrary to the expectation that for the $1/\sqrt{z}$ scaling of the Hamiltonian, the critical temperature diverges, as stated in \cite{Otsuki2013_2}. It turns out that when the average of the couplings remains finite and negative in the $z\rightarrow\infty$ limit, we are able to obtain spin order.

\begin{figure}[H]
    \centering
    \includegraphics[width=0.75\linewidth]{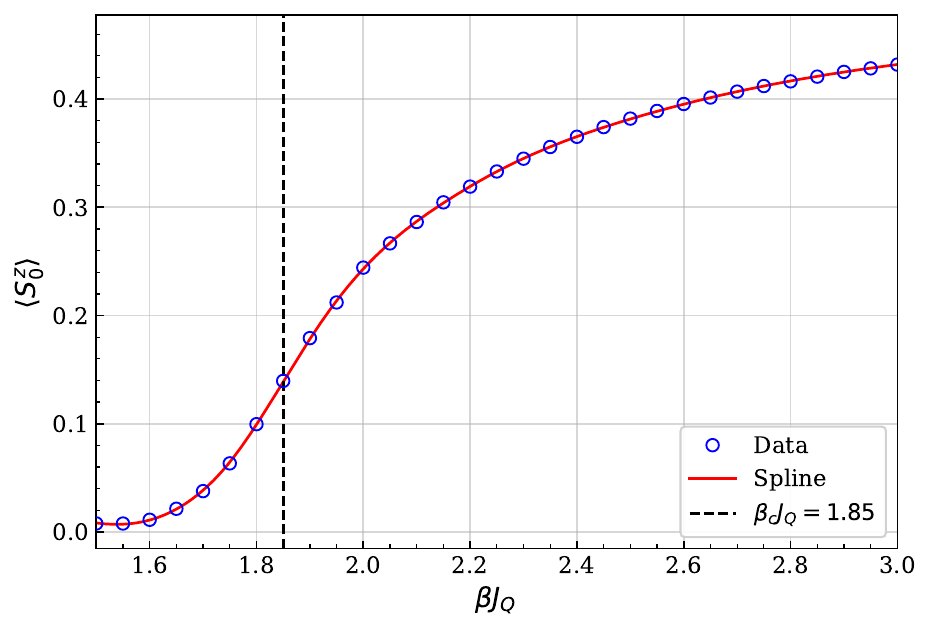}
    \caption{Dependence of the spin expectation value with a very small magnetic field $h_z=-0.001J_Q$ on temperature (blue points) and a numerically obtained spline function fitted to the points (red line).}
    \label{fig:M of beta}
\end{figure}
Now, we study the temperature dependence of the spin expectation value at near-zero external field ($h_z=-0.001 J_Q$) to determine the critical temperature at which order arises. The obtained data points are depicted in Fig. \ref{fig:M of beta}. Intriguingly, we do not see the magnetization disappearing but rather slowly decaying after a sudden drop. 
To extract the critical temperature, we compute a spline function numerically and find the inflection point by setting the second derivative to zero. 
The obtained estimate for the temperature is $\beta_c J_Q \approx1.85$. This is in contrast to the standard result from static mean-field theory \cite{Kittel1953}, which is given by 
\begin{equation}
    \beta_c J_Q = \frac{3J_Q}{|J_L|S(S+1)} = \frac{4}{\sqrt{6}}\approx1.63.
\end{equation}
This indicates that the dynamical fluctuations included in spinDMFT have a non-negligible effect on the critical temperature so that its value differs from what is reached by a static local mean-field treatment. This represents a promising route of further investigations.

It should be noted that spinDMFT leads to the same phase transition for an antiferromagnet on a bipartite lattice. If we take $J_L>0$ and assume the neighboring spins have the opposite expectation value to the spin at site $0$, the self-consistency \eqref{eq:mfaverages} is modified to
\begin{equation}
\overline{V^a(\tau_1)}  = -\JL \left<\mathbf{S}_0^a(\tau_1)\right>.
\end{equation}
We are left with a mathematically equivalent formulation to the ferromagnetic case if the minus sign is absorbed into $J_L$. Therefore all of the results remain the same, but now the computed spin expectation values can be interpreted as sublattice magnetization. The external magnetic field can be taken to be staggered, leading to identical field and temperature dependence and critical temperature. This underlines the limitation of single-site spinDMFT regarding the description of antiferromagnet order.

\section{Conclusion}\label{sec:conclusion}

In this work, we developed and tested a mean-field theory approach for the treatment of dense spin systems at finite temperature. It utilizes a time-dependent mean field to compute dynamical correlations in imaginary time. We argue that it is justified in the limit of an infinite coordination number in analogy to dynamical mean-field theory used for fermions. Hence we term our approach dynamical mean-field theory for spin systems (spinDMFT).

The key quantities for determining spin dynamics at finite temperatures are imaginary time-dependent autocorrelations and spin expectation values. We introduce an imaginary-time dependent mean field and argue that it is Gaussian distributed. The first and second moments of their probability distributions are connected to the autocorrelations and spin expectation values. This defines a self-consistency condition which we solve numerically by iteration of a stochastic approach.

The results of the method are gauged against exact results in small systems: a nearest-neighbor ferromagnet and antiferromagnet as well as a system with randomly sampled infinite-range couplings. We see good agreement in all cases at large temperatures, while at lower temperatures the results follow the correlations in the random-coupling system, deviating from the results on the other lattices. We also study the influence of an external magnetic field, once again obtaining the best agreement with the random-coupling system. The algorithm does not converge for low temperatures in the case of an antiferromagnet in an external magnetic field, presumably signaling that the
antiferromagnetic order requires an extension beyond the treatment of a single site. In the case of a ferromagnet, we see the the appearance of ferromagnetic order at low temperatures. Interestingly, the determined critical temperature is lower than the one from the standard static mean-field theory, indicating that the temporal fluctuations hamper ordering. All in all, spinDMFT produces quantitatively reliable results at lower temperatures only for the random-coupling system; in the other cases it reproduces the qualitative features of the correlations. For an
effective single-site theory we find this promising.


The mathematical underpinnings of the method remain an open problem. We did not provide a definite proof of the locality of the many-body problem in the $z\rightarrow\infty$ limit, nor of the exact form of the proposed self-consistency condition. The validity of this assumption certainly depends on the properties of the considered spin Hamiltonian, which we do not specify. A more thorough mathematical investigation could reveal the parameter regime in which the method is most effective. Yet, we believe that the advantageous comparisons with finite-size results are convincing enough to use spinDMFT also at finite temperature henceforth. Extensions of the method (discussed below) are expected to shine more light on these issues in the future.

There are multiple possibilities for extensions of spinDMFT at finite temperature. Including spin anisotropies, for instance, considering an XXZ instead of the isotropic Heisenberg Hamiltonian or adding Dzyaloshinskii–Moriya interactions, is relatively straightforward and amounts to a modification of the self-consistency conditions. Going beyond the single-site approximation by treating a larger spin cluster in a quantum manner has already been achieved at infinite temperature and is also straightforward to implement in our formalism. We expect it to improve agreement with finite-size results, in particular in the case of an antiferromagnetic system, where we could directly capture anti-aligned spins. 

A less computationally expensive alternative is considering multiple spins that are not directly coupled but interact through influencing each other's mean fields. One could also consider devising a modified self-consistency condition which takes into account the imaginary part of correlations. Another possible extension is directly computing the real-time evolution of spins. The constraints of our method would then force us to consider correlations of the mean-fields as a function of a complex parameter $t+\mathrm{i}\tau\in[0,\infty)\times[0,\beta]$ along a Keldysh contour, increasing the computational effort. However, the substantial benefit is the ability to simulate the evolution with time-dependent Hamiltonians and capture non-equilibrium effects. Also an analytic continuation would not be required to obtain real-time dynamics.

The establishment of spinDMFT at finite temperature opens up possibilities for new applications. SpinDMFT generally has enormous potential in the field of NMR. The approach is well suited to describe spin diffusion \cite{Suter1985, Graesser2026}, which is highly relevant in the research of dynamic nuclear polarization (DNP). Solid-state DNP is usually performed at cryogenic temperatures, where nuclear polarizations can become large enough that the infinite temperature assumption is no longer valid. Here, the presented finite-temperature version of spinDMFT can be very valuable. For extended solid state systems it is of great interest to capture phase transitions including dynamical correlations. Then, the dynamical structure factor becomes accessible and thereby the dispersion of magnetic excitations such as magnons, triplons and spinons as function of temperature. The renormalization and the broadening of their spectral lines would give access to the influence of temperature on their energy and on their life time. Hence, it is highly promising to pursue this
line of research further.

\section*{Acknowledgements}
We acknowledge useful discussions with Antonia J.\ Bock, Matthias Ernst, Piotr Kulik and Joachim Stolze.

\paragraph{Author contributions}
GSU formulated the scientific question and proposed the general approach. TG and PB worked
it out in detail. PB carried out the calculations, implemented the code, 
created the figures and wrote large parts of the manuscript. 
All authors interpreted the results and edited the manuscript.

\paragraph{Funding information}
PB thanks the Studienstiftung des deutschen Volkes for financial support by an individual grant.

\begin{appendix}
\numberwithin{equation}{section}

\section{Additional details of spinDMFT}

\subsection{The cavity method and Wick's theorem}
\label{ap:cavity_method}
In this section, we discuss the derivation of spinDMFT in more detail. The idea is to separate the degrees of freedom of spin $\vec{\mathbf{S}}_0$ from the ones of the remaining ensemble, which we refer to as cavity system. We present this derivation particularly for the spin autocorrelations, but the derivation for the spin expectation values and the partition function is analogous. The trace in Eq.~\eqref{eq:expval_autocorr_Z_after_trotterization} can be expanded according to
\begin{subequations}
\label{eq:cavityseparation}
\begin{align}
    g^{ab}(t,0) &= \sum_{m,n} \frac{(-1)^{m+n}}{m! n!} \int_{\tau}^{\beta} \mathrm{d}\tau_1 \dots \mathrm{d}\tau_m \int_{0}^{\tau} \mathrm{d}\tau_1^{'} \dots \mathrm{d}\tau_n^{'} \, f^{mn}(\{\tau\},\{\tau^{'}\}) \\
    f^{mn}(\{\tau\},\{\tau^{'}\}) &:= \sum_{a_1\dots a_m} \sum_{a_1^{'}\dots a_n^{'}} s^{mn}(\{\tau,a\},\{\tau^{'},a^{'}\}) w^{mn}(\{\tau,a\},\{\tau^{'},a^{'}\}) \\
    s^{mn}(\{\tau,a\},\{\tau^{'},a^{'}\}) &:= \sum_{m_0} \frac{Z_\text{C}}{Z} \bra{m_0} \mathcal{P}^{a_1\dots a_m}_{\tau_1\dots \tau_m} \left[\prod_{p=1}^{m} \oS_0^{a_p} \right]\Sa_0 \mathcal{P}^{a_1^{'}\dots a_n^{'}}_{\tau_1^{'} \dots \tau_n^{'}} \left[\prod_{q=1}^{n} \oS_0^{a_q^{'}} \right] \Sb_0 \ket{m_0} \\
    w^{mn}(\{\tau,a\},\{\tau^{'},a^{'}\}) &:= \sum_{W} \frac{1}{Z_\text{C}} \bra{W} \exp{-\beta\oW} \mathcal{T} \prod_{p=1}^{m} \tilde{\oV}_0^{a_p} (\tau_p) \prod_{q=1}^{n} \tilde{\oV}_0^{a_q^{'}} (\tau_q) \ket{W} \label{eq:wmn}.
\end{align}
\end{subequations}
where $Z_\text{C}$ is the partition function of the cavity system, $\ket{m_0}$ is a basis state of spin $\vec{\mathbf{S}}_0$ and $\ket{W}$ is a basis state of the cavity system. The permutation operator $\mathcal{P}$ orders the subsequent spin operators with indices $a$ in such a way that the attached times $\tau$ decrease from left to right. We used $\oVt:=\oV-\langle\oV\rangle$ to denote the operators shifted so that they have zero mean. Identifying $\exp{-\beta\oW}/Z_\text{C}$ as the thermal density operator of the cavity system, the expression in Eq.~\eqref{eq:wmn} can be interpreted as an expectation value computed in the cavity system. If the coordination number is large, removing a single spin $\vec{\mathbf{S}}_0$ from the ensemble produces an effect of order $1/z$ in local quantities around the removed site \cite{Georges1996}. Hence, in the limit $z\to\infty$, this expectation value is identical to an analogous expectation value in the full system,
\begin{align}
    w^{mn}(\{\tau,a\},\{\tau^{'},a^{'}\}) &= \left< \mathcal{T} \prod_{p=1}^{m} \tilde{\oV}_0^{a} (\tau_p) \prod_{q=1}^{n} \tilde{\oV}_0^{a_q} (\tau_q) \right>_{\text{C}} \overset{z\to\infty}{=} \left< \mathcal{T} \prod_{p=1}^{m} \tilde{\oV}_0^{a_p} (\tau_p) \prod_{q=1}^{n} \tilde{\oV}_0^{a_q} (\tau_q) \right>.
\end{align}
In principle, the spin and cavity degrees of freedom are separated in Eq.~\eqref{eq:cavityseparation}, but the evaluation of $w^{mn}$ is not feasible and one needs to take infinitely many terms into account. 
Using the fact that the fields $\oVt$ consist of many separate contributions we argue that they behave like classical variables because the relative norm of their commutators and vanish relative to the operators themselves \cite{Stanek2013}. In a second step, we argue again due to the many contributions to the fields that the central limit theorem holds for them so that Wick's theorem applies. Then $w^{mn}$ can be simplified so that Eq.~\eqref{eq:cavityseparation} can be evaluated by a single-site model with Gaussian mean fields. Hence, we use
\begin{subequations}
\label{eq:wickstheorem}
\begin{align}
    \langle \mathcal{T} \oVt_0^{a}(\tau_1) \oVt_0^{b}(\tau_2) \rangle &\approx \realpart\left\{ \langle \mathcal{T} \oVt_0^{a}(\tau_1) \oVt_0^{b}(\tau_2) \rangle \right\} \\
    \langle \mathcal{T} \oVt_0^{a}(\tau_1) \oVt_0^{b}(\tau_2) \oVt_0^{c}(\tau_3) \rangle &\approx 0 \\
    \begin{split}
    \langle \mathcal{T} \oVt_0^{a}(\tau_1) \oVt_0^{b}(\tau_2) \oVt_0^{c}(\tau_3) \oVt_0^{d}(\tau_4) \rangle &\approx \realpart\left\{\langle \mathcal{T}  \oVt_0^{a}(\tau_1) \oVt_0^{b}(\tau_2) \rangle\right\} \realpart\left\{\langle \mathcal{T} \oVt_0^{c}(\tau_3) \oVt_0^{d}(\tau_4) \rangle\right\}\\
    & + \realpart\left\{\langle \mathcal{T} \oVt_0^{a}(\tau_1) \oVt_0^{c}(\tau_3) \rangle\right\} \realpart\left\{\langle \mathcal{T} \oVt_0^{b}(\tau_2) \oVt_0^{d}(\tau_4) \rangle\right\} \\
    & + \realpart\left\{\langle \mathcal{T} \oVt_0^{a}(\tau_1) \oVt_0^{d}(\tau_4) \rangle\right\} \realpart\left\{\langle \mathcal{T}  \oVt_0^{b}(\tau_2) \oVt_0^{c}(\tau_3) \rangle\right\},
    \end{split}
\end{align}
\end{subequations}
and analogous approximations for higher-order terms. A rigorous proof that this approximation becomes exact for $z\to\infty$ is lacking so that it remains an assumption. However, analogous to infinite-temperature spinDMFT, there is a strong argument for Wick's theorem: If $z$ is large, the environment fields $\oV_0$ consist of a large number of spins, which due to the scaling of the couplings $J\propto1/\sqrt{z}$ are only weakly correlated. Therefore, the conditions of the central limit theorem are fulfilled, which implies that the field is Gaussian distributed and its moments separate according to Wick's theorem. Taking the real part in Eq.~\eqref{eq:wickstheorem} corresponds to an additional assumption, which will later ensure that the mean fields are real-valued. At infinite temperature, this is irrelevant, since all correlations remain purely real. At finite temperature, correlations can become complex so that taking the real value does make a difference. An extension to complex mean fields, where the real part is not taken, is not studied in this work.

\subsection{Implementation}\label{ap:spindmft implementation}
The numerical implementation of spinDMFT consists of two main steps: sampling of the mean fields and computation of correlations with the drawn fields. First, we use initial correlations and spin expectation values to build the covariance matrix \eqref{eq:cov matrix} and the average vector \eqref{eq:average vector}. To simplify the calculations, we can use the symmetry of the considered problem. For instance, when the external magnetic field vanishes, $\vec{B}=0$, the Hamiltonian \eqref{eq:full Hamiltonian} becomes spin isotropic. As discussed in section \ref{sec:spin correlations}, in this case the diagonal correlations are all equal and the off-diagonal correlations and the spin expectation values vanish. We can enforce this symmetry in the implementation, setting $\mathcal{M}_{\tau_1\tau_2}^{xx}=\mathcal{M}_{\tau_1\tau_2}^{yy}=\mathcal{M}_{\tau_1\tau_2}^{zz}$ and $\mathcal{M}_{\tau_1\tau_2}^{ab}=0$ for $a\neq b$. Then, only one spin correlation, e.g. $g^{xx}$ needs to be calculated to define the whole covariance matrix. The sampling is simplified because the covariance matrix is block-diagonal and the $x,\,y$ and $z$ component can be drawn separately. In the case with a magnetic field $\vec{B}=B\vec{e}_z$, one can similarly use the invariance under rotations around the $z$ axis for a reduction of the number of relevant correlations.

In order to draw the mean fields from the probability distribution \eqref{eq:prob distr}, we need to diagonalize the covariance matrix. After drawing the random mean fields in the diagonal basis, they are transformed back to the original basis. Since the matrix is time-translation-invariant $\mathcal{M}_{\tau_1\tau_2}^{ab}=\mathcal{M}_{\tau_1-\tau_2,0}^{ab}$, diagonalization is most easily achieved through a Fourier transform to Matsubara frequencies
\begin{equation}
    \mathcal{M}^{ab}(\omega_n) = \frac{1}{\sqrt{\beta}}\int_0^\beta d\tau\,e^{-\mathrm{i}\omega_n\tau}\mathcal{M}_{\tau,0}^{ab}.
\end{equation}
Then, the covariance matrix becomes block-diagonal (the blocks are 3$\times$3 and correspond to spin indices). The self-consistency condition takes the form
\begin{equation}
    M^{ab}(\omega_n) = J_Q^2\left(\mathrm{Re}\,g^{ab}(\omega_n)-\left<\mathbf{S}_0^a\right>\left<\mathbf{S}_0^b\right>\sqrt{\beta}\delta_{n,0}\right),
\end{equation}
with $\omega_n=2\pi n/\beta$ being the bosonic Matsubara frequencies. After drawing the mean fields $\vec{V}(\omega_n)$, the fields in the imaginary time domain are obtained by a Fourier back transform. Using the frequency representation is way more efficient than direct diagonalization of the covariance matrix. The frequency approach scales as $L\,\mathrm{log}\,L$ with the number of time points, compared to the $L^3$ scaling of diagonalization.

For drawing samples, we use the preconditioned Crank-Nicolson (pCN) algorithm \cite{Cotter2013}. To make the procedure clear, let us denote the sampled traces as
\begin{subequations}
\begin{align}
    C^{ab}(\tau)[\vec{\mathcal{V}}] &= \trace{ \mathcal{T}e^{-\int^\beta_\tau\mathbf{H}_{\mathrm{mf}}(\tau')[\vec{\mathcal{V}}]d\tau'}\mathbf{S}^a_0e^{-\int_0^\tau \mathbf{H}_{\mathrm{mf}}(\tau')[\vec{\mathcal{V}}]d\tau'}\mathbf{S}^b_0 };\\
    Z_0[\vec{\mathcal{V}}] &= \trace{ \mathcal{T}e^{-\int^\beta_0\mathbf{H}_{\mathrm{mf}}(\tau')[\vec{\mathcal{V}}]d\tau'} }.
\end{align}
\end{subequations}
Then, the sample average in the computation of correlations can be rewritten as
\begin{equation}
    g^{ab}(\tau) = \frac{1}{Z_0}\int\mathcal{D \vec{V}}p[\vec{\mathcal{V}}]C^{ab}(\tau)[\vec{\mathcal{V}}] = \int\mathcal{D \vec{V}}\frac{p[\vec{\mathcal{V}}]Z_0[\vec{\mathcal{V}}]}{Z_0} \frac{C^{ab}(\tau)[\vec{\mathcal{V}}]}{Z_0[\vec{\mathcal{V}}]}.
\end{equation}
This means that the correlations can be computed as an average of $C^{ab}(\tau)[\vec{\mathcal{V}}]/Z_0[\vec{\mathcal{V}}]$ with a temperature-dependent probability distribution $p_\beta[\vec{\mathcal{V}}]=p[\vec{\mathcal{V}}]Z_0[\vec{\mathcal{V}}]/Z_0$. It is a useful point of view for two reasons. First, it avoids potential overflows which might appear for large sample sizes when computing $Z_0$ and $C^{ab}(\tau)$ separately. Second, at low temperatures we noticed that $p_\beta[\vec{\mathcal{V}}]$ peaks very far from the original distribution $p[\vec{\mathcal{V}}]$. This can be understood by realizing, that $Z_0[\vec{\mathcal{V}}]\propto e^{\beta|\vec{\mathcal{V}}|}$ so its effect is to shift the weight of $p[\vec{\mathcal{V}}]$ away from the mean. This implies that sampling the mean fields from $p_\beta[\vec{\mathcal{V}}]$ instead of $p[\vec{\mathcal{V}}]$ drastically reduces the variance at low temperatures (in some cases even by a factor $10^3$), requiring orders of magnitude less samples for accurate results. However, $p_\beta[\vec{\mathcal{V}}]$ includes $Z_0$, which itself would need to be determined by sampling. Therefore, we use Markov chain Monte Carlo, which only uses ratios of probabilities, eliminating the need for the normalization factor. The final ingredient is 
pCN, which is designed for sampling probability distributions of the form $P(X)=P_0(X)\mathrm{exp}(-\Phi(X))$, where $P_0(X)$ is a Gaussian distribution and $\Phi(X)$ is a real-valued function. This is exactly our problem with $P_0=p[\vec{\mathcal{V}}]$ and $\Phi=-\mathrm{log}(Z_0[\vec{\mathcal{V}}]/Z_0)$. Then, the pCN 
prescription is: given an initial guess $\vec{\mathcal{V}}_0$, generate a new one using
\begin{equation}
    \vec{\mathcal{V}}_{n+1} = \sqrt{1-\gamma^2}\vec{\mathcal{V}}_{n} + \gamma \vec{\mathcal{V}}',
\end{equation}
where $\vec{\mathcal{V}}'$ is drawn from the Gaussian $p[\vec{\mathcal{V}}]$ and $\gamma\in[0,1]$ is a real parameter which can be tuned to maximize efficiency. Then, accept the new configuration with probability
\begin{equation}
    \alpha(\vec{\mathcal{V}}_{n+1},\vec{\mathcal{V}}_{n}) = \mathrm{min}\left(1, Z_0[\vec{\mathcal{V}}_{n+1}]/Z_0[\vec{\mathcal{V}}_{n}]\right).
\end{equation}
After accepting/rejecting the configuration, accumulate $C^{ab}(\tau)[\vec{\mathcal{V}}]/Z_0[\vec{\mathcal{V}}]$ into $g^{ab}(\tau)$. All of the above considerations apply to the computation of the spin expectation values $\langle\vec{\mathbf{S}_0}\rangle$ as well.

After drawing a given configuration of the mean fields $\vec{\mathcal{V}}=(\vec{V}(0),\vec{V}(\Delta\tau),\ldots,\vec{V}(\beta))^T$, we use it to compute evolution operators $U(\tau,0)$ for $\tau\in(0, \Delta\tau,\ldots,\beta)$ as in Eq. \eqref{eq:partition_mf}. To this end, we express them as
\begin{equation}
    \mathbf{U}(\tau,0) = \mathbf{U}(\tau,\tau-\Delta\tau)\mathbf{U}(\tau-\Delta\tau,\tau-2\Delta\tau)\ldots \mathbf{U}(\Delta\tau,0).
\end{equation}
Then, the evolution over time steps $\Delta\tau$ can be efficiently computed with commutator-free exponential time propagation (CFET) \cite{Alvermann2011}. 
We truncate at fourth order in $\Delta\tau$, leading to the following approximate formula
\begin{align}
    \mathbf{U}_{\mathrm{CFET}4}(\tau_l,\tau_{l-1}) =& \mathrm{exp}\left(\frac{11}{40}\mathbf{A}_1+\frac{20}{87}\mathbf{A}_2+\frac{7}{50}\mathbf{A}_3\right)\mathrm{exp}\left(\frac{9}{20}\mathbf{A}_1-\frac{7}{25}\mathbf{A}_3\right)\notag\\
    &\times\mathrm{exp}\left(\frac{11}{40}\mathbf{A}_1-\frac{20}{87}\mathbf{A}_2+\frac{7}{50}\mathbf{A}_3\right), 
\end{align}
where
\begin{equation}
    \mathbf{A}_j =(2j-1)\frac{\Delta\tau}{2}\left(\mathbf{H}_{\mathrm{mf}}(\tau_l)-(-1)^j\mathbf{H}_{\mathrm{mf}}(\tau_{l-1})\right).
\end{equation}
The matrix exponentials are efficiently computed using the Pauli matrix exponential formula
\begin{equation}
    \mathrm{exp}(-\vec{a}\cdot\vec{\sigma}) = \mathrm{cosh}(|\vec{a}|)\mathbb{1} - \frac{\mathrm{sinh}(|\vec{a}|)}{|\vec{a}|}\vec{a}\cdot\vec{\sigma},
\end{equation}
since the mean-field Hamiltonian can always be expressed as a sum of Pauli matrices.


\section{Chebyshev expansion technique}
\label{ap:numerical methods}

The Chebyshev expansion technique \cite{Tal-Ezer1984, Kosloff1994, Weisse2006} is a method for approximating the action of the evolution operator on the states by expanding it in Chebyshev polynomials of the Hamiltonian. The Chebyshev polynomials are defined recursively by
\begin{equation}
    T_0(x)=1,\quad T_1(x)=x,\quad T_{n}(x)=2xT_{n-1}(x)-T_{n-2}(x).
\end{equation}
To expand the evolution operator, the Hamiltonian must be rescaled such that the spectrum is contained in the interval $[-1,1]$. This can be done by defining $\mathbf{H}'=(\mathbf{H}-b)/a$, with $a,b$ depending on the smallest and largest eigenvalue of the Hamiltonian $E_{\mathrm{min}},E_{\mathrm{max}}$ as $a=(E_{\mathrm{max}}-E_{\mathrm{min}})/2$ and $b=(E_{\mathrm{max}}+E_{\mathrm{min}})/2$. The extrema of the spectrum can be obtained straightforwardly with the Lanczos algorithm. Then, the evolution operator can be expressed as
\begin{align}
    U(\tau) =& \sum_{n=0}^\infty \alpha_n(\tau)T_n(\mathbf{H}');\\
    \alpha_n(\tau) =& (2-\delta_{n,0})e^{b\tau}I_n(a\tau),
\end{align}
with $I_n$ being the $n$-th modified Bessel function of the first kind. In practice, it is more efficient to consider the action of the operator on states. Given an initial state $|\psi_0\rangle$, its time evolution can be computed as
\begin{equation}
    |\psi(\tau)\rangle=\sum_{n=0}^\infty\alpha_n(\tau)|\psi_n\rangle,
\end{equation}
with the states $|\psi_n\rangle$ given by a recursion
\begin{equation}
    |\psi_1\rangle = \mathbf{H}'|\psi_0\rangle,\quad |\psi_n\rangle = 2\mathbf{H}'|\psi_{n-1}\rangle-|\psi_{n-2}\rangle. 
\end{equation}
When implementing the method, the infinite sum in the expansion has to be truncated at some cut-off order $N_{\mathrm{c}}$. The truncation error scales as
\begin{equation}
    \epsilon_{\mathrm{CET}} \propto \mathcal{O}\left(\frac{\sqrt{a\tau}}{N_{\mathrm{c}}}\right)^{N_{\mathrm{c}}}.
\end{equation}
From this we see that the cut-off order required to minimize the error grows with the bandwidth of the Hamiltonian and with the time the evolution is carried out for. For that reason, it is profitable to conduct the time evolution over small timesteps, as it minimizes the required cut-off order.

In the case of a spin-$\frac{1}{2}$ Heisenberg Hamiltonian, the action of the Hamiltonian on states can be implemented using bitwise operations to reduce computation time. The main advantage of the algorithm is its memory efficiency. When computing evolution with this approach, only the action of the Hamiltonian is required, not its matrix form. Therefore, it is enough to save states in memory instead of matrices.

\section{Comparison to spin glass physics}\label{ap:grempel}

\begin{figure}[H]
    \centering
    \includegraphics[width=0.75\textwidth]{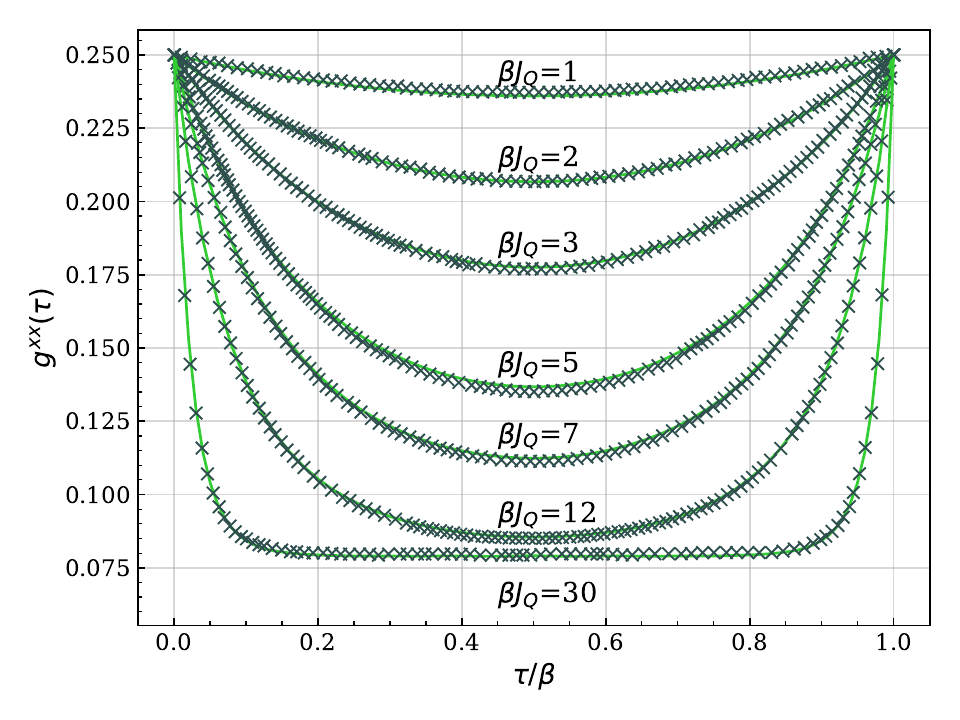}
    \caption{Comparison of correlations obtained in spinDMFT (green line) and in a spin glass mean-field theory  developed in Ref.\ \cite{Grempel1998} (gray crosses) at different temperatures (specified above the curves). In the spin glass data, there are errors due to the graphical extraction of the published data; they are roughly of the size of the crosses. The error bars of spinDMFT are smaller than the line widths.}
    \label{fig:spindmft vs grempel}
\end{figure}

In Ref.\ \cite{Grempel1998}, Grempel and Rozenberg implemented a mean-field theory approach for spin glass systems based on a self-consistency condition derived previously \cite{Bray1980}. The mathematical structure of the approach is very similar to our method, especially in the isotropic case, despite a completely different derivation (relying on the properties of the spin glass systems). In Fig.\ \ref{fig:spindmft vs grempel}, we compare the results of spinDMFT without an external magnetic field with the data depicted in a plot from Ref.\ \cite{Grempel1998}. We see that both methods yield almost the same correlations in this case. This further emphasizes the connection of spinDMFT to spin glass physics. We stress, however, that the advantage of our approach is a more general derivation, applicable for instance to systems without a special distribution of couplings, in a magnetic field, with spin anisotropies and to larger spin clusters.

\section{Real-time evolution and finite-size effects}
\label{ap:real time}

In this appendix, we obtain
real-time data from the imaginary-time correlations computed from spinDMFT and compare them to finite-size results. We also consider the relevance of finite-size effects, which differ between real- and imaginary-time correlations. To compute the real-time dynamics, we employ the maximum entropy method \cite{Jarrell1996} to perform stochastic analytic continuation of imaginary-time correlations to real time. We used the publicly available ana\_cont implementation \cite{Kaufmann2023, Bergeron2016} and focus on the case without an external magnetic field.
\begin{figure}[H]
    \centering
    \includegraphics[width=\linewidth]{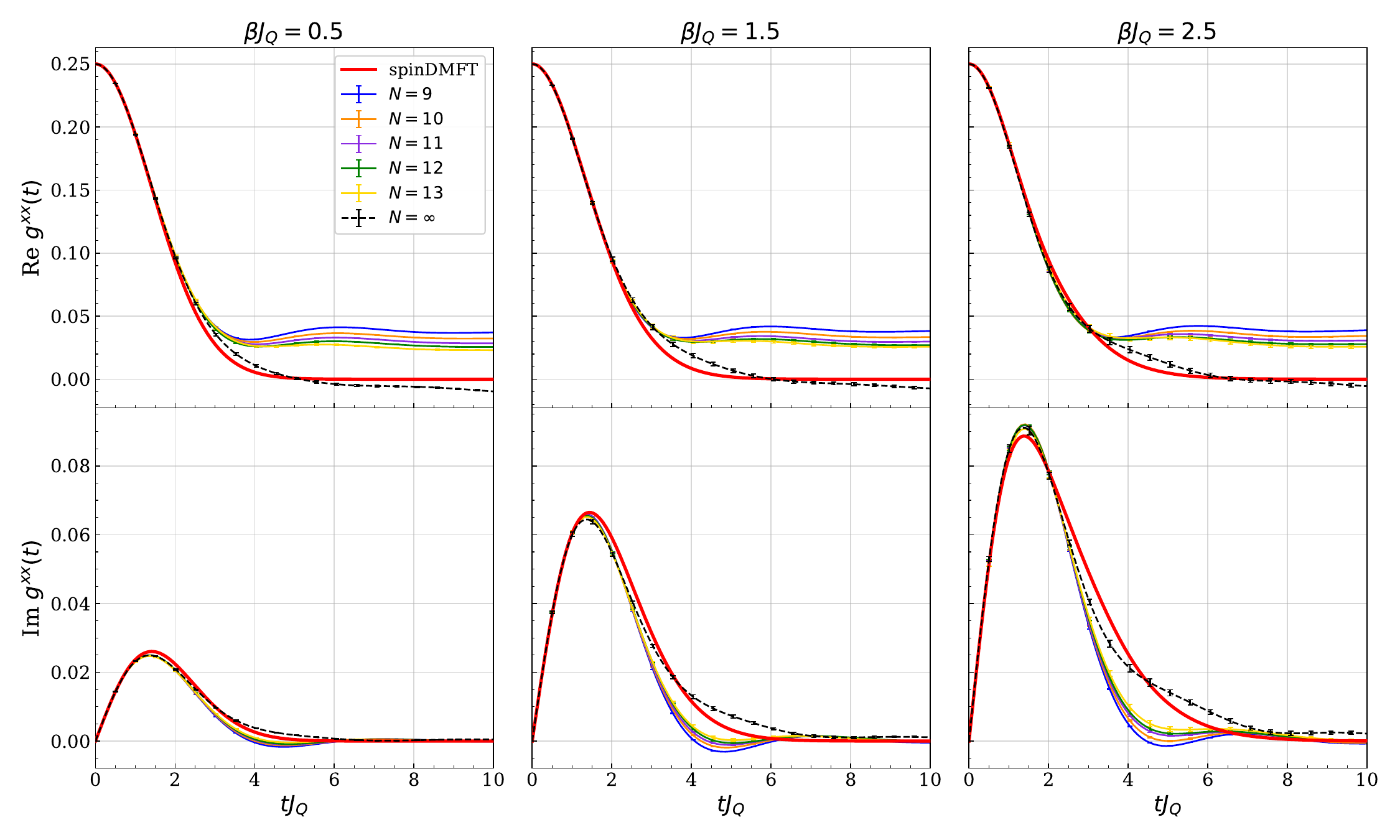}
    \caption{Comparison between real-time correlations in a random-coupling system obtained from spinDMFT by analytic continuation (thick red line), and obtained from CET on a finite-size system of varying size  (thin lines). The black line corresponds to the $1/N$ extrapolation of the finite-size data. The three columns depict data at different values of temperature, growing from left to right. The real part is plotted in the top row and the imaginary part in the bottom row.}
    \label{fig:realtime_random}
\end{figure}

\begin{figure}[H]
    \centering
    \includegraphics[width=\linewidth]{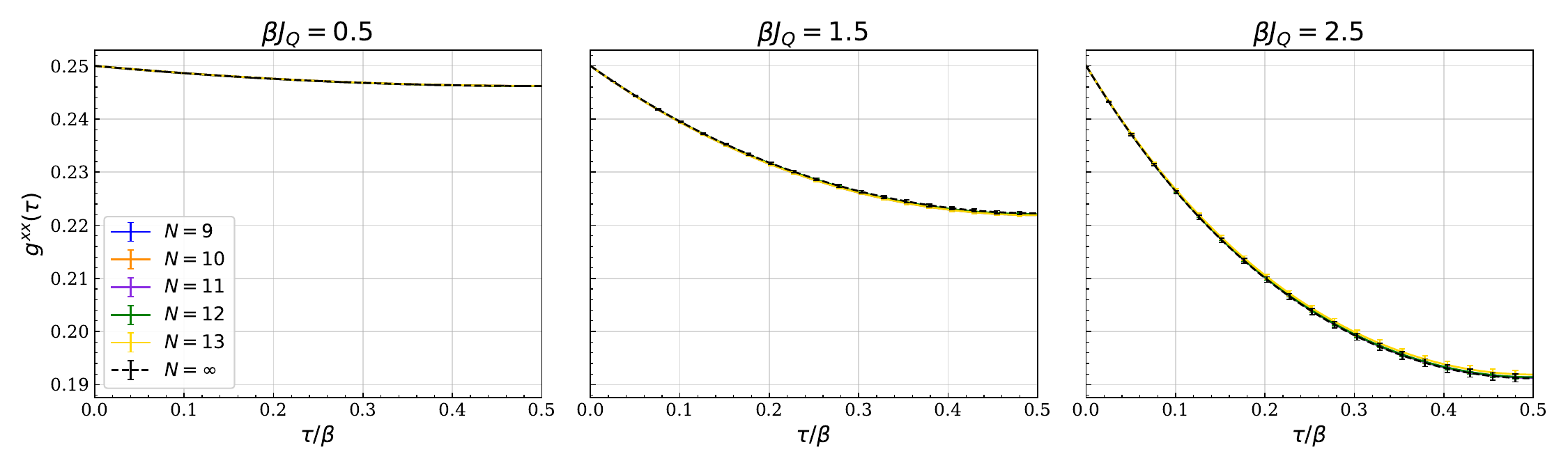}
    \caption{Comparison between imaginary-time correlations obtained from CET in a random-coupling system of different sizes including $1/N$ extrapolation. One can see, that all finite-size deviations are below the numerical error. The three columns depict data at different values of temperature, growing from left to right.}
    \label{fig:random_finitesize}
\end{figure}

In Fig. \ref{fig:realtime_random}, we depict the comparison of the results to real-time correlations obtained in a random-coupling system with CET at three values of the temperature. We plot finite-size data at various system sizes as well as an $1/N$ extrapolation with $N\rightarrow\infty$. One can see, that finite-size effects play a 
role for real-time correlations, especially for long times. However, the extrapolation data agree very well with spinDMFT for short times. At intermediate times we note noticeable deviations for smaller temperature, but they can likely be attributed to finite-size effects. For comparison with imaginary-time dynamics, see Fig. \ref{fig:random_finitesize}, where we plot the imaginary-time correlations at various system sizes together with an extrapolation. All of the curves are almost identical, indicating that finite-size effects play a much smaller role in imaginary time.

\begin{figure}[H]
    \centering
    \includegraphics[width=\linewidth]{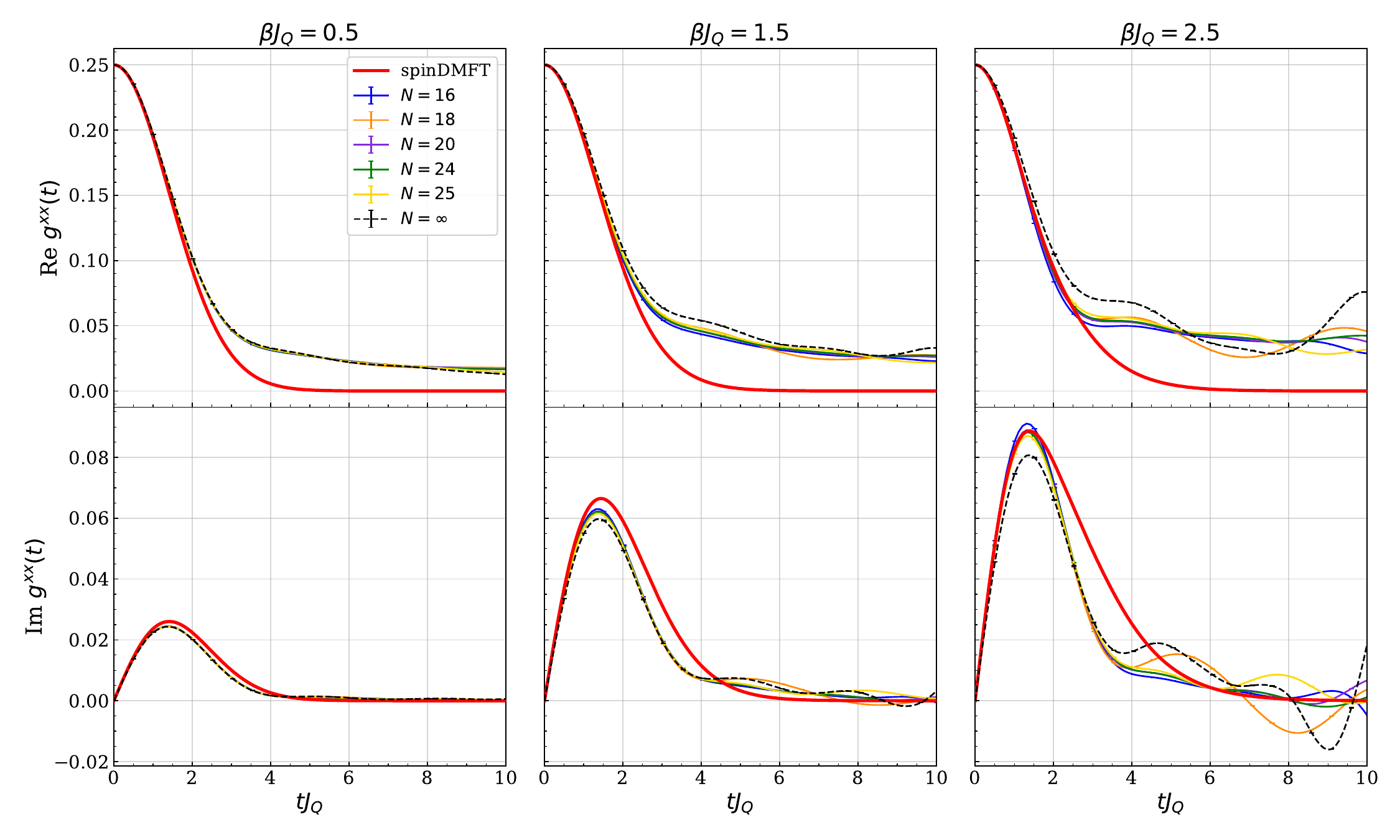}
    \caption{Same as Figure~\ref{fig:realtime_random}, but for a ferromagnetic system on a square lattice.}
    \label{fig:realtime_square}
\end{figure}

In Fig. \ref{fig:realtime_square}, we depict a comparison of the results to real-time correlations obtained in a ferromagnetic system on a square lattice with periodic boundary conditions. Once again, we compute the correlations with CET and perform a $1/N$ extrapolation. We note, that for this small-$z$ system, the finite-size effects are much more pronounced. This likely makes the extrapolation less trustworthy. Nonetheless, at very high temperatures the agreement of spinDMFT with the finite-size data is good for short times. At lower temperatures, the finite-size effects are much stronger and do not allow for a definite comparison. However it is clear that spinDMFT does not agree quantitatively with the data from the ferromagnet.

\end{appendix}

\bibliography{SciPost_Example_BiBTeX_File.bib}

\end{document}